\documentclass[twocolumn,showpacs,preprintnumbers,amsmath,amssymb]{revtex4}

\usepackage{graphicx}
\usepackage{dcolumn}
\usepackage{bm}%

\newtheorem{theorem}{Theorem}

\newtheorem{definition}[theorem]{Definition}

\newenvironment{proof}[1][Proof]{\textbf{#1.} }{\ \rule{0.5em}{0.5em}}

\newcommand{\C}{\mathbb C}

\newcommand{\R}{\mathbb R}

\def\la{\label}
\def\bt{\begin{thm}}
\def\et{\end{thm}}

\def\bl{\begin{lem}}
\def\el{\end{lem}}

\def\bd{\begin{defi}}
\def\ed{\end{defi}}

\def\bc{\begin{cor}}
\def\ec{\end{cor}}

\def\bp{\begin{proof}}
\def\ep{\end{proof}}

\def\br{\begin{rem}}
\def\er{\end{rem}}

\newtheorem{thm}{Theorem}[section]
\newtheorem{lem}{Lemma}[section]
\newtheorem{defi}{Definition}[section]

\newtheorem{rem}{Remark}[section]
\newtheorem{cor}{Corollary}[section]

\numberwithin{equation}{section}
\numberwithin{theorem}{section}
\numberwithin{example}{section}

\begin{document}

\title{Dynamic Phase Transitions in PVT Systems}

\thanks{The work was supported in part by the
Office of Naval Research and by the National Science Foundation.}
\author{Tian Ma}
\affiliation{Department of Mathematics, Sichuan University,
Chengdu, P. R. China}%

\author{Shouhong Wang}
 \homepage{http://www.indiana.edu/~fluid}
\affiliation{Department of Mathematics,
Indiana University, Bloomington, IN 47405}%
\date{\today}

\begin{abstract}
The main objective of this article are two-fold. First, we introduce some general principles on phase transition dynamics, including a new dynamic transition classification scheme, and a Ginzburg-Landau theory for modeling equilibrium phase transitions. Second, apply the general principles and the recently developed dynamic transition theory to study dynamic phase transitions of PVT systems. In particular, we establish a new time-dependent  Ginzburg-Landau model, whose dynamic transition analysis is carried out. It is worth pointing out that the new dynamic transition theory, along with the dynamic classification scheme and new time-dependent Ginzburg Landau models for equilibrium phase transitions can be used in other phase transition problems, including e.g. the ferromagnetism and superfluidity, which will be reported elsewhere. In addition,  the analysis for the PVT system in this article leads to a few physical predications, which are otherwise unclear from the physical point of view.
\end{abstract}

\maketitle

\section{Introduction}
\label{sc1}

The main objectives of this article are 

\begin{itemize}

\item[1)] to introduce a new dynamic phase transition classification scheme for nonlinear systems, 

\item[2)] to present a new time-dependent Ginzburg-Landau model for general equilibrium phase transitions, and 

\item[3)] to study the dynamic phase transitions for PVT systems. 

\end{itemize}

As we know, there are several classification schemes for phase transitions. The one used most often is the classical Ehrenfest classification scheme, which is based on the order of differentiability of the energy functional.  The new dynamic phase transition classification scheme we propose here 
is based  on the  dynamic transition theory developed recently by the authors 
\cite{chinese-book, b-book}, which has been used to study numerous problems in sciences and engineering, including in particular problems in classical and geophysical fluid dynamics, biology and chemistry, and phase transitions. 
The new scheme introduced in this article  classifies phase transitions into three categories:  Type-I, Type-II and Type-III, corresponding respectively to the continuous, the jump and mixed transitions in the mathematical dynamic transition theory. 

We shall see from the applications to PVT systems that this new classification scheme is more transparent. In addition, it can be used to identify high-order (in the Ehrenfest sense)  transitions,  which are usually hard to derive both theoretically and experimentally. 

Second, based on the le Ch\^atelier principle and some general characteristics of pseudo-gradient systems, we introduce a general principle, leading to a unified approach to derive Ginzburg-Landau type of time-dependent models for equilibrium phase transitions. In view of this new dynamic modeling and the dynamic transition theory, we shall see that  the states after transition often include not only equilibria, but also some transient states, which are physically important as well.

Third, with the aforementioned principles in our disposal, we derive a new dynamic model for PVT system, whose dynamic transition can be explicitly obtained, leading to some specific physical predictions. 

As we know, a $PVT$ system is a system composed of one type  of molecules, and  the interaction between molecules is governed by the van
der Waals law. The molecules generally have a repulsive core and
a short-range attraction region outside the core. Such systems
have a number of phases: gas, liquid and solid, and a solid can
appear in a few phases. The most typical example of a $PVT$ system is
water. In general, the phase transitions of $PVT$ systems mainly
refer to the gas-liquid, gas-solid and liquid-solid phase
transitions. These transitions are all first order in the Ehrenfest sense (i.e., 
discontinuous) and are accompanied by a latent heat and a change
in density. 

A $PT$-phase diagram of a typical $PVT$ system is schematically
illustrated by Figure \ref{f8.1}, where point $A$ is the triple point
at which the gas, liquid, and solid phases can coexist. Point $C$
is the Andrews critical point at which the gas-liquid coexistence
curve terminates \cite{reichl}. The fact that the gas-liquid coexistence curve
has a critical point means that we can go continuously from a gaseous  state 
to a liquid state without ever meeting an observable phase transition,
if we choose the right path. The critical point in the
liquid-solid coexistence curve has never been observed. It implies
that we must go through a first-order phase transition in going
from the liquid to the solid state.

\begin{figure}[hbt]
  \centering
  \includegraphics[height=.4\hsize]{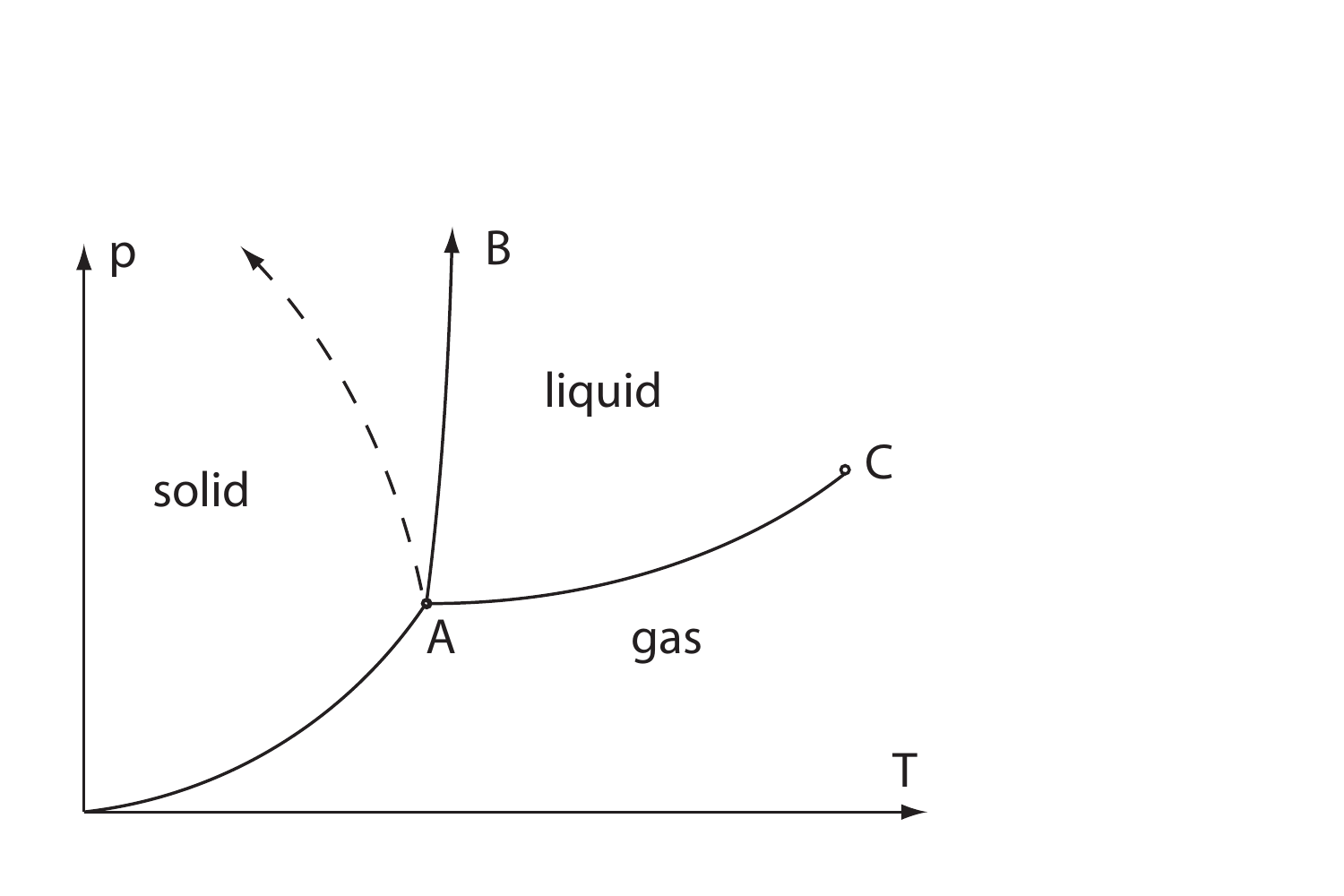}
  \caption{Coexistence curves of a typical $PVT$ system:
$A$ is the triple point, $C$ is the critical point, and the dashed
cure is a melting line with negative slope.}\la{f8.1}
 \end{figure}

The phase diagram in the $PV$-plane is given by Figure \ref{f8.2}, 
where the dashed curves represent lines of constant temperature.
In the region of coexistence of phases, the isotherms (dashed
lines) are always flat, indicating that in these regions the
change in volume (density) occurs for constant pressure $p$ and temperature $T$.
\begin{figure}[hbt]
  \centering
  \includegraphics[height=.5\hsize]{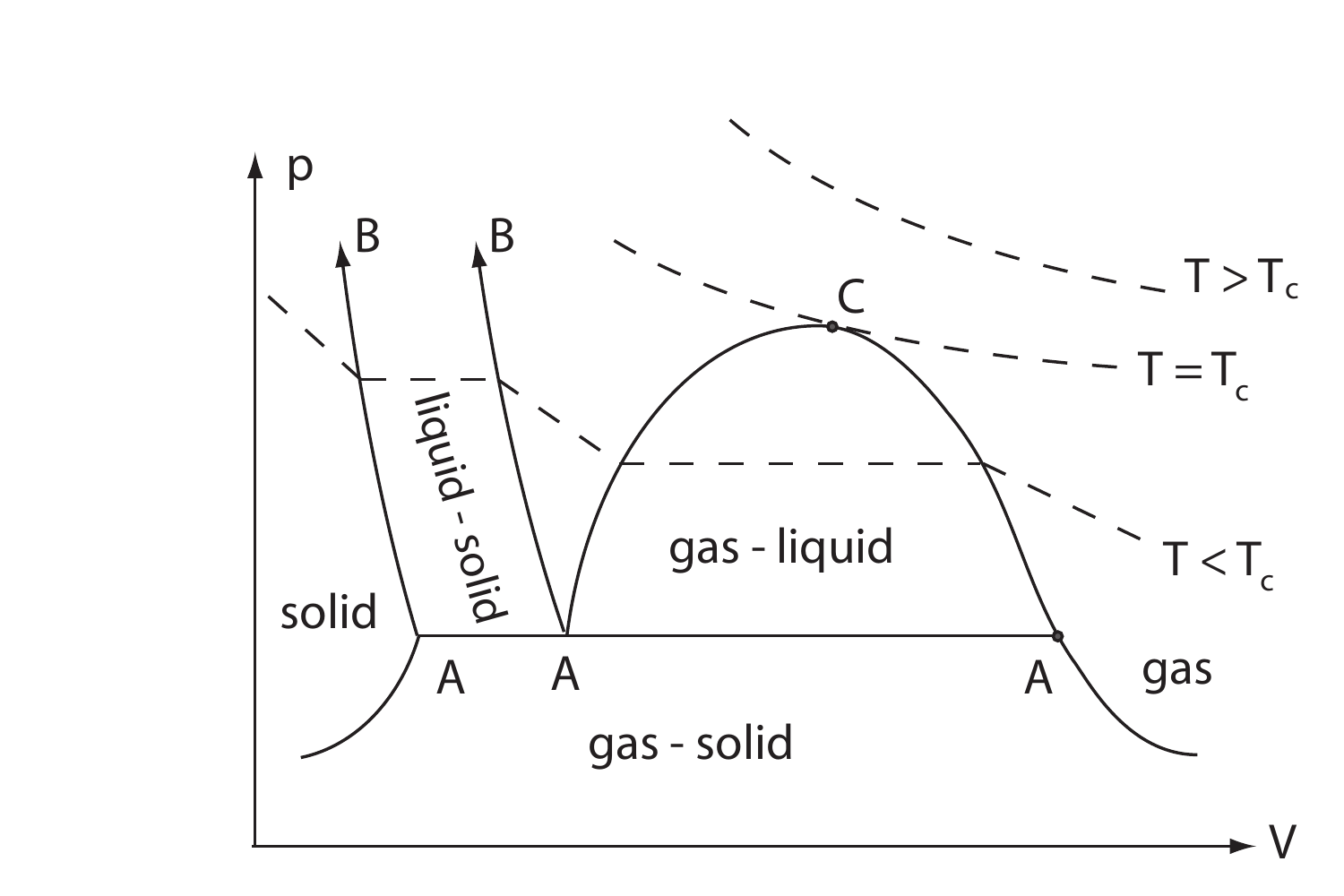}
  \caption{$PV$-phase diagram, the dashed line represent
isotherms.}\la{f8.2}
 \end{figure}

One of the main objectives of this article is to study the dynamic phase transitions beyond the Andrews point $C$, and to make some physical predications.

For this purpose, first, by both mathematical and physical considerations, a proper form of the Gibbs free energy functional is given, and then the time-dependent Ginzburg-Landau type of model for PVT systems is derived naturally using the general principle as mentioned above. 

Then, using the dynamical transition theory, we  derive a system of critical parameter equations to determine the Andrews point $C=(T_C,p_C)$ given by  (\ref{8.19}).
Then we show that the gas-liquid transition can be either first-order, second order or third order, as demonstrated in Physical Conclusion V.1, 
if we go from a gas to a liquid by a path without passing
through the gas-liquid coexistence curve.

In addition, in the gas-solid and liquid-solid
transitions,  there also exist metastable states. For the gas-solid
case, the metastable states correspond to the superheated solid
and supercooled liquid, and for the liquid-solid case, the
metastable states correspond to the superheated solid and
supercooled liquid; see Physical Conclusion V.2.

This article is organized as follows. The new dynamic classification scheme and the Ginzburg-Landau model for equilibrium phase transitions are given in Section II. Section III introduces a time-dependent model for PVT systems, which is analyzed in Section IV. Section V provides physical consequences of the analysis and some new  physical  
predictions.The abstract dynamical transition theory is recapitulated in the Appendix A for convenience.

\section{General Principles of Phase Transition Dynamics}
In this section,  we introduce a new phase dynamic transition classification scheme 
to classify phase transitions into three categories: Type-I, Type-II and Type-III, corresponding mathematically  continuous,  jump mixed transitions, respectively. 

Then  a new time-dependent Ginzburg-Landau theory for modeling  equilibrium phase transitions in statistical physics is derived based on the le Ch\^atelier principle and some mathematical insights on pseudo-gradient systems. 

\subsection{Dynamic Classification Scheme}
In sciences, nonlinear dissipative systems are generally governed
by differential equations, which can be expressed in the following
abstract form 
\begin{equation}
\left.
\begin{aligned} 
&\frac{du}{dt}=L_{\lambda}u+G(u,\lambda ),\\
&u(0)=\varphi , 
\end{aligned} \right.\label{7.1}
\end{equation}
where $u:[0,\infty )\rightarrow X$ is the unknown function,
$\lambda\in \R^N$ $(N\geq 1)$ is  the control parameter, $X$  and $X_1$ are two Banach
spaces with $X_1\subset X$ being a  dense and compact inclusion,
$L_{\lambda}=-A+B_{\lambda}$ and $G(\cdot ,\lambda
):X_1\rightarrow X$ are $C^r(r\geq 0)$ mappings depending
continuously on $\lambda$,  $L_{\lambda}:X_1\rightarrow X$ is a
sectorial operator, and
\begin{equation}
\left.
\begin{aligned} &A:X_1\rightarrow X  && \text{a\ linear\
homeomorphism},\\
&B_{\lambda}:X_1\rightarrow X&& \text{a\ linear\ compact\
operator}.
\end{aligned}
\right.\label{7.2}
\end{equation}

In following, we introduce some basic and universal concepts in
nonlinear sciences.  

First, a  state of the system (\ref{7.1}) at $\lambda$ is usually referred to as a compact invariant set $\Sigma_{\lambda}$. In many applications, 
$\Sigma_{\lambda}$ is a singular point or a periodic orbit. A state $\Sigma_{\lambda}$
of (\ref{7.1}) is stable if $\Sigma_{\lambda}$ is an attractor,
otherwise $\Sigma_{\lambda}$ is called unstable.

Second,  we say that the system (\ref{7.1}) has a
phase transition from a state $\Sigma_{\lambda}$ at $\lambda
=\lambda_0$ if $\Sigma_{\lambda}$ is stable on $\lambda <\lambda_0$
(or on $\lambda >\lambda_0$) and is unstable on $\lambda
>\lambda_0$ (or on $\lambda <\lambda_0$).  The critical parameter $\lambda_0$ is
called a critical point.
 In other words, the
phase transition corresponds to an exchange of stable states.

The concept of phase transition originates from the statistical
physics and thermodynamics. In physics and chemistry, "phase"
means the homogeneous part in a heterogeneous system. However,
here the so called phase means the stable state in the systems of
nonlinear sciences including physics, chemistry, biology, ecology,
economics, fluid dynamics and geophysical fluid dynamics, etc.
Hence, here the content of phase transition has been endowed with
more general significance. In fact, the phase transition dynamics
introduced  here can be applied to a wide variety of topics involving
the universal critical phenomena of state changes in nature in a
unified mathematical viewpoint and manner.

Third, if the system (\ref{7.1}) possesses the gradient-type structure, then the phase transitions are called equilibrium phase transition; otherwise they are called the non-equilibrium phase transitions.

Fourth, classically, there are several ways to classify phase transitions. The one most used  is the Ehrenfest classification scheme, which groups phase transitions based on the degree of non-analyticity involved.  First order phase transitions are  also called discontinuous, and  higher order phase transitions $(n>1)$ are called continuous.

Here we introduce the following notion of  dynamic classification scheme: 

\begin{definition}
Let $\lambda_0\in \R^N$ be a critical point of (\ref{7.1}), and (\ref{7.1}) undergo a transition from state $\Sigma^1_{\lambda}$ to $\Sigma^2_{\lambda}$. There are three types of phase transitions for  (\ref{7.1}) at $\lambda =\lambda_0$, depending on their  dynamic properties: continuous, jump, and mixed as given in Theorem~\ref{t5.1}, which are called Type-I, Type-II and Type-III respectively.
\end{definition}

The main characteristics  of Type-II phase transitions is that there is a gap between $\Sigma^1_{\lambda}$ and $\Sigma^2_{\lambda}$ at the critical point $\lambda_0$. In thermodynamics,  the metastable states correspond in general to the super-heated or super-cooled states, which have been found in many physical phenomena. In particular,
Type-II phase transitions are always accompanied with the latent heat to occur.

In a Type-I phase transition, two states $\Sigma^1_{\lambda}$ and 
$\Sigma^2_{\lambda}$ meet  at $\lambda_0$, i.e., the system undergoes 
a continuous transition from $\Sigma^1_{\lambda}$ to
$\Sigma^2_{\lambda}$.

In a Type-III phase transition, there are at least two different stable states $\Sigma^{\lambda}_2$ and $\Sigma^{\lambda}_3$ at $\lambda_0$, and system undergoes a continuous transition to $\Sigma^{\lambda}_2$ or a jump transition  to $\Sigma^{\lambda}_3$, depending on the  fluctuations.

It is clear that a Type-II phase transition of gradient-type
systems must be discontinuous or the zero order because there is a
gap between $\Sigma^1_{\lambda}$ and $\Sigma^j_{\lambda}$  $(2\leq j\leq
K)$. For a Type-I phase transition, the energy  is continuous, and consequently, it is an $n$-th order transition in the Ehrenfest sense  for some $n\geq 2$.
A Type-III phase transition is indefinite, for the transition from
$\Sigma^1_{\lambda}$ to $\Sigma^2_{\lambda}$ it may be continuous, i.e.,
$dF^-/d\lambda =dF^+_2/d\lambda$, and for the transitions from
$\Sigma^1_{\lambda}$ to $\Sigma^3_{\lambda}$ it may be discontinuous:
$dF^-/d\lambda\neq dF^+_3/d\lambda$ at $\lambda =\lambda_0$.

Finally, we recall the physical meaning of the derivatives of
$F(u,\lambda )$ on $\lambda$ in statistical physics. In
thermodynamics, the parameter $\lambda$ generally stands for
temperature $T$ and pressure $p$
$$\lambda =(T,p).$$
The energy functional $F(u,T,p)$ is the thermodynamic potential, and
$u$ is the order parameter. The first order derivatives are
\begin{align*}
&-\frac{\partial F}{\partial T}=S&& \text{the\ entropy},\\
&\frac{\partial F}{\partial p}=V&& \text{the\ phase\ volume},
\end{align*}
and the second order derivatives are
\begin{align*}
&-T\frac{\partial^2F}{\partial T^2}=C_p&& \text{the\ heat\ capacity\
in\ constant\ pressure},\\
&-\frac{1}{V}\frac{\partial^2F}{\partial p^2}=\kappa && \text{the\
compression\ coefficient},\\
&\frac{1}{V}\frac{\partial^2F}{\partial T\partial p}=\alpha && \text{the\ thermal\ expansion\ coefficient},
\end{align*}
for the $PVT$ system.

Thus, for the first order phase transition, the discontinuity of
$\partial F/\partial p$ at critical point $\lambda_0=(T_0,p_0)$
implies the discontinuity of phase volume:
\begin{equation}
\Delta V=V^2-V^1=\frac{\partial F^+}{\partial p}-\frac{\partial
F^-}{\partial p},\label{7.10}
\end{equation}
and the discontinuity of $\partial F/\partial T$ implies that
there is a gap between both phase entropies:
\begin{equation}
\Delta S=S^2-S^1=-\frac{\partial F^+}{\partial T}+\frac{\partial
F^-}{\partial T}\label{7.11}
\end{equation}

Physically, $\Delta S$ is a non-measurable quantity, hence, we
always use latent heat $\Delta H=T\Delta S$ to determine the first
order phase transition, and $\Delta H$ stand for the absorbing
heat for transition from phase $\Sigma^1_{\lambda}$ to phase
$\Sigma^2_{\lambda}$. In the two phase coexistence situation,
$\Delta H$ and $\Delta V$ are related by the Clausius-Clapeyron
equation  $$\frac{dp}{dT}=\frac{\Delta H}{T\Delta V}.$$

For second order phase transitions the variance of heat capacity
(or specific heat), compression coefficient, and the thermal
expansion coefficient at critical value are important static
properties in equilibrium phase transition, which are measurable
physical quantities.

\subsection{New Ginzburg-Landau Models for Equilibrium Phase Transitions}
\label{s7.2.2}
In this subsection, we introduce the time-dependent Ginzburg-Landau model for equilibrium phase transitions. 

We start with thermodynamic potentials and the Ginzburg-Landau free energy. As we know, four thermodynamic potentials-- internal energy, the enthalpy, the Helmholtz free energy and the Gibbs free energy--are useful in the chemical thermodynamics of reactions and non-cyclic processes. 

Consider a thermal system, its order parameter $u$ changes in
$\Omega\subset \R^n$ $(1\leq n\leq 3)$. In this situation,  the free energy of this system is of the form
\begin{equation}
{\mathcal{H}}(u,\lambda
)= {\mathcal{H}}_0 
 +\int_{\Omega}\Big[\frac{1}{2}\sum^m_{i=1}\mu_i|\nabla
u_i|^2  +g(u,\nabla u,\lambda )\Big]dx\label{7.28}
\end{equation}
where $N\geq 3$ is an integer,
$u=(u_1,\cdots,u_m),\mu_i=\mu_i(\lambda )>0$, and $g(u,\nabla
u,\lambda )$ is a $C^r(r\geq 2)$ function of $(u,\nabla u)$ with
the Taylor expansion \begin{equation} g(u,\nabla u,\lambda
)=\sum\alpha_{ijk}u_iD_ju_k+\sum^N_{|I|=1}\alpha_Iu^I+o(|u|^N)-fX,\label{7.29}
\end{equation}
where $I=(i_1,\cdots,i_m),i_k\geq 0$ are integer,
$|I|=\sum^m_{k=1}i_k$, the coefficients $\alpha_{ijk}$ and
$\alpha_I$ continuously depend on $\lambda$, which are determined
by the concrete physical problem, $u^I=u^{i_1}_1\cdots u^{im}_m$
and $fX$ the generalized work.

Thus, the study of thermal equilibrium phase transition for the
static situation is referred to the steady state bifurcation of
the system of elliptic equations
$$\left.
\begin{aligned}
&\frac{\delta}{\delta u}{\mathcal{H}}(u,\lambda )=0,\\
&\frac{\partial u}{\partial n}|_{\partial\Omega}=0,\ \ \ \
(\text{or}\ u|_{\partial\Omega}=0),
\end{aligned}
\right.$$ where $\delta /\delta u$  is the variational derivative.

A thermal system is controled by some parameter $\lambda$. When
$\lambda$ is for from the critical point $\lambda_0$ the system
lies on a stable equilibrium state $\Sigma_1$, and when $\lambda$
reaches or exceeds $\lambda_0$ the state $\Sigma_1$ becomes unstable, and 
meanwhile the system will undergo a transition  from $\Sigma_1$ to another stable
state $\Sigma_2$. The basic principle is that there often exists  fluctuations in the system leading  to a deviation from the equilibrium states, and  the phase transition process is a
dynamical behavior, which should be described by a time-dependent
equation.

To derive a general time-dependent model, first we recall that  the classical  
le Ch\^atelier  principle amounts to saying that 
for a stable  equilibrium state of a system $\Sigma$, when the system deviates from
$\Sigma$ by a small perturbation or fluctuation, there will be a
resuming force to retore this system to return to the stable state
$\Sigma$.
Second, we know that  a stable equilibrium state of a thermal system must
be the minimal value point of the thermodynamic potential. 

By the mathematical characterization of gradient systems and the le Ch\^atelier principle, for a system with
thermodynamic potential ${\mathcal{H}}(u,\lambda )$, the governing
equations are essentially determined by the functional
${\mathcal{H}}(u,\lambda )$.
When the order parameters $(u_1,\cdots,u_m)$ are nonconserved
variables, i.e., the integers
$$\int_{\Omega}u_i(x,t)dx=a_i(t)\neq\text{constant}.$$
then the time-dependent equations are given by
\begin{equation}
\left.
\begin{aligned} 
&\frac{\partial u_i}{\partial
t}=-\beta_i\frac{\delta}{\delta u_i}{\mathcal{H}}(u,\lambda
)+\Phi_i(u,\nabla u,\lambda ),\\
&\frac{\partial u}{\partial n}|_{\partial\Omega}=0\ \ \ \
(\text{or}\ u|_{\partial\Omega}=0),\\
&u(x,0)=\varphi (x),
\end{aligned}
\right.\label{7.30}
\end{equation}
for any $1 \le i \le m$, 
where $\delta /\delta u_i$ are the variational derivative,
$\beta_i>0$ and $\Phi_i$ satisfy
\begin{equation}
\int_{\Omega}\sum_i\Phi_i\frac{\delta}{\delta
u_i}{\mathcal{H}}(u,\lambda )dx=0.\label{7.31}
\end{equation}
The condition (\ref{7.31})  is  required by
the Le Ch\^atelier principle. In the concrete problem, the terms
$\Phi_i$ can be determined by physical laws and (\ref{7.31}).

When the order parameters are the number density and the system
has no material exchange with the external, then $u_j$  $(1\leq j\leq
m)$ are conserved, i.e.,
\begin{equation}
\int_{\Omega}u_j(x,t)dx=\text{constant}.\label{7.32}
\end{equation}
This conservation law requires a continuous equation
\begin{equation}
\frac{\partial u_j}{\partial t}=-\nabla\cdot J_j(u,\lambda
),\label{7.33}
\end{equation}
where $J_j(u,\lambda )$ is the flux of component $u_j$. In
addition, $J_j$ satisfy
\begin{equation}
J_j=-k_j\nabla (\mu_j-\sum_{i\neq j}\mu_i),\label{7.34}
\end{equation}
where $\mu_l$ is the chemical potential of component $u_l$, 
\begin{equation}
\mu_j-\sum_{i\neq j}\mu_i=\frac{\delta}{\delta
u_j}{\mathcal{H}}(u,\lambda )-\phi_j(u,\nabla u,\lambda
), \label{7.35}
\end{equation}
and  $\phi_j(u,\lambda )$ is a function depending on the other
components $u_i$ $(i\neq j)$. When $m=1$, i.e., the system consists of
two components $A$ and $B$, this term $\phi_j=0$. Thus, from
(\ref{7.33})-(\ref{7.35}) we obtain the dynamical equations as
follows
\begin{equation}
\left.
\begin{aligned} &\frac{\partial u_j}{\partial
t}=\beta_j\Delta\left[\frac{\delta}{\delta
u_j}{\mathcal{H}}(u,\lambda )-\phi_j(u,\nabla u,\lambda )\right],\\
&\frac{\partial u}{\partial n}|_{\partial\Omega}=0,\ \ \ \
\frac{\partial\Delta u}{\partial n}|_{\partial\Omega}=0,\\
&u(x,0)=\varphi (x),
\end{aligned}
\right.\label{7.36}
\end{equation}
for $1 \le j \le m$, 
where $\beta_j>0$ are constants, $\phi_j$ satisfy
\begin{equation}
\int_{\Omega}\sum_j\Delta\phi_j\cdot\frac{\delta}{\delta
u_j}{\mathcal{H}}(u,\lambda )dx=0.\label{7.37}
\end{equation}

If the order parameters $(u_1,\cdots,u_k)$ are coupled to the
conserved variables $(u_{k+1},\cdots,u_m)$, then the dynamical
equations are
\begin{equation}
\left.
\begin{aligned} 
&\frac{\partial u_i}{\partial t}
   =-\beta_i\frac{\delta}{\delta u_i}{\mathcal{H}}(u,\lambda)+\Phi_i(u,\nabla u,\lambda ),\\
& \frac{\partial u_j}{\partial t}
  =\beta_j\Delta\left[\frac{\delta}{\delta u_j}{\mathcal{H}}(u,\lambda )
    -\phi_j(u,\nabla u,\lambda )\right],\\
&\frac{\partial u_i}{\partial n}|_{\partial\Omega}=0\ \ \ \
(\text{or}\ u_i|_{\partial\Omega}=0),\\
&\frac{\partial u_j}{\partial n}|_{\partial\Omega}=0,\ \ \ \
\frac{\partial\Delta u_j}{\partial n}|_{\partial\Omega}=0,\\
&u(x,0)=\varphi (x).
\end{aligned}
\right.\label{7.38}
\end{equation}
for $1 \le i \le k$  and $k+1 \le j \le m$.

The model (\ref{7.38}) gives a general form of the governing
equations to thermodynamic phase transitions.  Hence, the dynamics of
equilibrium phase transition in statistic physics is based on the new Ginzburg-Landau formulation  (\ref{7.38}).

Physically, the initial value condition $u(0)=\varphi$ in
(\ref{7.38}) stands for the fluctuation of system or perturbation
from the external. Hence, $\varphi$ is generally small. However,
we can not exclude the possibility of a bigger noise $\varphi$.

From conditions (\ref{7.31}) and (\ref{7.37}) it follows that a
steady state solution $u_0$ of (\ref{7.38}) satisfies
\begin{equation}
\left.
\begin{aligned} 
&  \Phi_i(u_0,\nabla u_0,\lambda )=0 && \forall 1\leq
i\leq k,\\
& \Delta\phi_j(u_0,\nabla u_0,\lambda )=0 && \forall k+1\leq j\leq m.
\end{aligned}
\right.\label{7.39}
\end{equation}
Hence a stable equilibrium state must reach the
minimal value of thermodynamic potential. In fact, $u_0$ fulfills
\begin{equation}
\left.
\begin{aligned} 
&
\beta_i\frac{\delta}{\delta u_i}{\mathcal{H}}(u_0,\lambda )
   -\Phi_i(u_0,\nabla u_0,\lambda)=0,\\
&\beta_j\Delta\frac{\delta}{\delta u_j}{\mathcal{H}}(u_0,\lambda)
   -\Delta\phi_j(u_0,\nabla u_0,\lambda )=0,\\
&\frac{\partial u_i}{\partial n}|_{\partial\Omega}=0\ \ \ \
(\text{or}\ u_i|_{\partial\Omega}=0),\\
\\
&\frac{\partial u_j}{\partial n}|_{\partial\Omega}=0,\ \ \ \
\frac{\partial\Delta u_j}{\partial n}|_{\partial\Omega}=0,
\end{aligned}
\right.\label{7.40}
\end{equation}
for $1 \le i \le k$ and $k+1 \le j \le m$.
Multiplying $\Phi_i(u_0,\nabla u_0,\lambda )$ and
$\phi_j(u_0,\nabla u_0,\lambda )$ on the first and the second
equations of (\ref{7.40}) respectively, and integrating them, then
we infer from (\ref{7.31}) and (\ref{7.37}) that
\begin{eqnarray*}
&&\int_{\Omega}\sum_i\Phi^2_i(u_0,\nabla u_0,\lambda )dx=0,\\
&&\int_{\Omega}\sum_j|\nabla\phi_j(u_0,\nabla u_0,\lambda
)|^2dx=0,
\end{eqnarray*}
which imply that (\ref{7.39}) holds  true.

\section{Time-Dependent Model for $PVT$ Systems}
\la{s8.1.1}
In this section, we use the general principles derived in the last section to derive a 
time-dependent Ginzburg-Landau model for PVT systems, which will be used to carry out dynamic transition analysis and physical predications in the next section.

\subsection{van der Waals equations and Gibbs energy}


The classical and the simplest equation of state which can
exhibit many of the essential features of the gas-liquid phase
transition is the van der Waals equation. It reads
\begin{equation}
v^3-\left(b+\frac{RT}{p}\right)v^2+\frac{a}{p}v-\frac{ab}{p}=0,\label{8.1}
\end{equation}
where $v$ is the molar  volume, $p$ is  the pressure, $T$ is the
temperature, $R$ is the universal gas constant, $b$ is  the revised
constant of inherent volume, and  $a$  is  the revised constant of attractive
force between molecules. If we adopt the molar density $\rho =1/v$
to replace $v$ in (\ref{8.1}), then the van der Waals equation
becomes
\begin{equation}
-(bp+RT)\rho +a\rho^2-ab\rho^3+p=0.\label{8.2}
\end{equation}

Now, we shall apply thermodynamic potentials to investigate the
phase transitions of $PVT$ systems, and we shall  see later that the
van der Waals equation can be derived as a Euler-Langrange equation for the minimizers of the Gibbs free energy for $PVT$ systems at gaseous states.

Consider an isothermal-isopiestic  process. The thermodynamic
potential is taken to be the Gibbs free energy. In this case, the
order parameters are the molar density $\rho$ and the  entropy $S$, and
the control parameters are the pressure $p$ and temperature $T$. 
The general form of the  Gibbs free energy 
for $PVT$ systems  is given as 
\begin{align}
G(\rho ,S,T,p)   = &
 \int_{\Omega}\Big[\frac{1}{2}\mu_1|\nabla\rho
|^2+\frac{1}{2}\mu_2|\nabla S|^2+g(\rho ,S,T,p)
 \nonumber \\
 &  -ST-\alpha (\rho ,T,p)\rho \Big]dx, \label{8.3}
 \end{align}
where $g$ and $\alpha$ are differentiable with respect to  $\rho$ and $S$,
$\Omega\subset \R^3$ is the container, and $\alpha p$ is the
mechanical coupling term in the Gibbs free energy, which can be
expressed by
\begin{equation}
\alpha (\rho ,T,p)p=\rho p-\frac{1}{2}b\rho^2p, \label{8.4}
\end{equation}
where $b=b(T,p)$ depends continuously on $T$ and $p$.

Based on both the  physical and mathematical considerations, 
we take the Taylor expansion of $g(\rho ,S,T,p)$ on $\rho$ and $S$ as follows
\begin{equation}
g=\frac{1}{2}\alpha_1\rho^2+\frac{1}{2}\beta_S^2+\beta_2S\rho^2-\frac{1}{3}\alpha_2\rho^3+\frac{1}{4}\alpha_3\rho^4,\label{8.5}
\end{equation}
where $\alpha_i$  $(1\leq i\leq 3)$, $\beta_1$  and $\beta_2$ depend
continuously on $T$ and $p$, and
\begin{equation}
\left.
\begin{aligned} 
& \alpha_i=\alpha_i(T,p)>0 &&i=2,3,\\
& \beta_1=\beta_1(T,p)>0.
\end{aligned}
\right.\label{8.6}
\end{equation}

\subsection{Dynamical equations for PVT systems}
 In a $PVT$ system, the order parameter is $u=(\rho ,S)$,
$$\rho =\rho_1-\rho_0,\ \ \ \ S=S_1-S_0,$$
where $\rho_i$ and $S_i(i=0,1)$ represent the density and entropy, 
$\rho_0,S_0$ are reference points. Hence the
conjugate variables of $\rho$ and $S$ are the pressure $p$ and
the temperature $T$. 
Thus, by the standard model (\ref{7.30}), we derive from (\ref{8.3})-(\ref{8.5}) 
the following general form of the time-dependent equations
governing a $PVT$ system: 
\begin{equation}
\left.
\begin{aligned} 
&\frac{\partial\rho}{\partial
t}=\mu_1\Delta\rho -(\alpha_1+bp)\rho
+\alpha_2\rho^2\\
&\qquad -\alpha_3\rho^3-2\beta_2\rho S+p,\\
&\frac{\partial S}{\partial t}=\mu_2\Delta
S-\beta_1S-\beta_2\rho^2+T.
\end{aligned}
\right.\label{8.7}
\end{equation}

Although the domain $\Omega$ depends on $T$ and $p$, we can still
take the Neumann boundary condition
\begin{equation}
\frac{\partial\rho}{\partial n}\left|_{\partial\Omega}=0,\ \ \ \
\frac{\partial S}{\partial
n}\right|_{\partial\Omega}=0.\label{8.8}
\end{equation}

An important special case for  $PVT$ systems is that the
pressure and temperature functions are homogeneous in $\Omega$. Thus we can
assume that $\rho$ and $S$ are independent of $x\in\Omega$,  and 
 the free energy (\ref{8.3}) with (\ref{8.4}) and
(\ref{8.5}) can be expressed as
\begin{align}
 G(\rho,S,T,p) 
\nonumber
 = &\frac{1}{2}\alpha_1\rho^2+\frac{1}{2}\beta_1S^2+\beta_2S\rho^2-\frac{1}{3}\alpha_2\rho^3\\
 &+
\frac{1}{4}\alpha_3\rho^4
+\frac{1}{2}b\rho^2p-\rho p-ST.  \label{8.9}
\end{align}

From (\ref{8.9}) we get the dynamical equations as
\begin{equation}
\left.
\begin{aligned} &\frac{d\rho}{dt}=-(\alpha_1+bp)\rho
+\alpha_2\rho^2\\
& \qquad -\alpha_3\rho^3-2\beta_2S\rho +p,\\
&\frac{dS}{dt}=-\beta_1S-\beta_2\rho^2+T.
\end{aligned}
\right.\label{8.10}
\end{equation}
Because $\beta_1>0$ for all $T$ and $p$, we can replace the second
equation of (\ref{8.10}) by 
\begin{equation}
S=\beta^{-1}_1(T-\beta_2\rho^2).
\end{equation}
Then, (\ref{8.10}) are equivalent to the following equation
\begin{align} \nonumber
\frac{d\rho}{dt}= & -(\alpha_1+bp+2\beta^{-1}_1\beta_2T)\rho
+\alpha_2\rho^2\\
& -(\alpha_3-2\beta^2_2\beta^{-1}_1)\rho^3+p.\label{8.11}
\end{align}

It is clear that if  $\alpha_1=0, 2\beta^{-1}_1\beta_2=R,
\alpha_2=a, (\alpha_3-2\beta^2_2\beta^{-1}_1)=ab$,   then the steady
state equation of (\ref{8.11}) is referred to the van der Waals
equation.

\section{Phase Transition Dynamics for PVT Systems}
In this section we use (\ref{8.11}) to discuss dynamical properties of
transitions for $PVT$ systems, and remark that similar results can also derived using the 
more general form of the time-dependent model (\ref{8.7}).

Let $\rho_0$ be a steady state
solution of (\ref{8.11}). We take the transformation
$$\rho =\rho_0+\rho^{\prime}.$$
Then equation (\ref{8.11}) becomes (drop the prime)
\begin{equation}
\frac{d\rho}{dt}=\lambda\rho +a_2\rho^2-a_3\rho^3.\label{8.12}
\end{equation}
where
\begin{align*}
&\lambda
=2\alpha_2\rho_0-3a_3\rho^2_0-\alpha_1-bp-2\beta_2\beta^{-1}_1T,\\
&a_2=\alpha_2-3a_3\rho_0,\\
&a_3=\alpha_3-2\beta^2_2\beta^{-1}_1.
\end{align*}

In   the $PT$-plane, near a non-triple point $A=(T^*,p^*)$ (see Figure
\ref{f8.1}),  the critical parameter equation
$$\lambda =\lambda (T,p)=0 \qquad  \text{in}\ \ \ \ |T-T^*|<\delta
,\ \ \ \ |p-p^*|<\delta$$ 
for some $\delta >0$,
defines  a continuous function $T=\phi (p)$, such that
\begin{equation}
\lambda
\left\{
\begin{aligned} 
& <0  &&  \text{ if }  T>\phi (p),\\
& =0  &&  \text{ if } T=\phi (p),\\
& >0  &&  \text{ if } T<\phi (p).
\end{aligned}
\right.\label{8.13}
\end{equation}

The main result in this section is the following dynamic transition theorem for PVT systems.

\bt\la{t8.1} Let $T_0=\phi (p_0)$ and $a_3>0$. Then the
system (\ref{8.12}) has a transition at $(T,p)=(T_0,p_0)$, and the following assertions hold true:
\begin{enumerate}

\item  If the
coefficient $a_2=a_2(T,p)$ in (\ref{8.12}) is zero at $(T_0,p_0)$,
i.e., $a_2(T_0,p_0)=0$, then the transition is of Type-I, as
schematically shown in Figure \ref{f8.3} (a) and (b). 

\item  If
$a_2(T_0,p_0)\neq 0$,  then the transition  is of Type-III (i.e., the mixed type), and the following assertions hold true:

\begin{enumerate}
\item   There are two transition solutions near $(T_0,p_0)$ as
\begin{equation}
\rho^{\pm}(T,p)=\frac{1}{2a_3}\left(a_2\pm\sqrt{a^2_2+4a_3\lambda}\right).\la{rhopm}
\end{equation}

\item  There is a saddle-node bifurcation at $(T_1,p_1)$,  where 
$T_1>T_0$  and $ p_1<p_0$ if  $\phi^{\prime}(p_0)>0$,  and $p_0>p_1$
if  $\phi^{\prime}(p_0)<0$.

\item  For $\phi^{\prime}(p_0)>0$, when $a_2(T_0,p_0)>0$ the
transition diagrams are illustrated by Figure \ref{f8.4} (a)-(b), 
where $\rho^+$ is stable for all $(T,p)$ near $(T_0,p_0)$, and
$\rho =0$ is stable, $\rho^-$ a saddle for $T_0<T<T_1$ and
$p_1<p<p_0$, and $\rho^-$ is stable, $\rho =0$ a saddle for
$T<T_0, p>p_0$;

\item  When $a_2(T_0,p_0)<0$, the transition diagrams are illustrated by
Figure \ref{f8.5}(a)-(b), where $\rho^-$ is stable for all $(T,p)$
near $(T_0,p_0)$, and $\rho =0$ is stable, $\rho^+$ a saddle for
$T_0<T<T_1, p_1<p<p_0$, and $\rho^+$ is stable, $\rho =0$ a saddle
for $T<T_0, p_0<p$.
\end{enumerate}
\end{enumerate}
\et
\begin{figure}[hbt]
  \centering
  \includegraphics[width=0.35\textwidth]{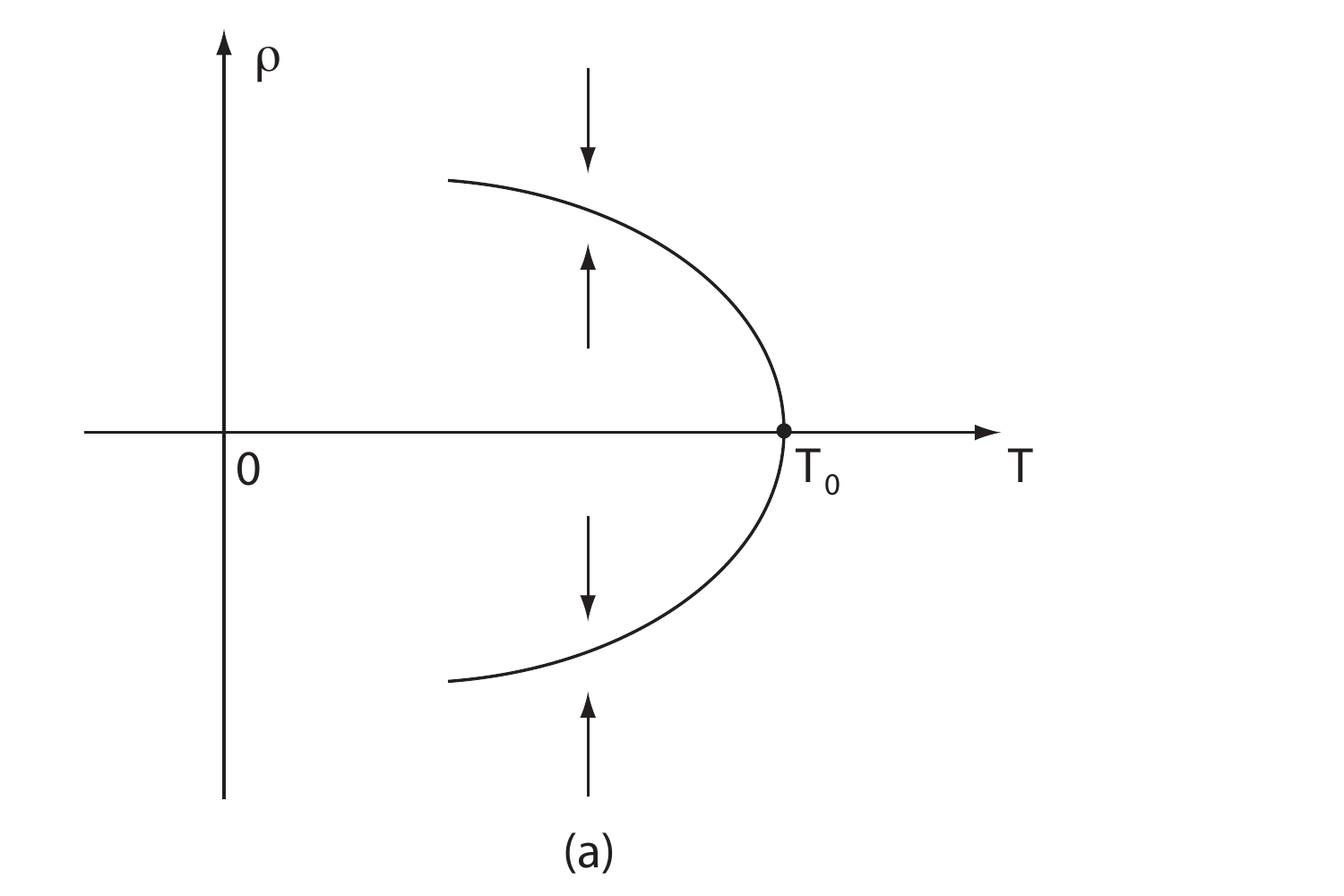}
  \includegraphics[width=0.35\textwidth]{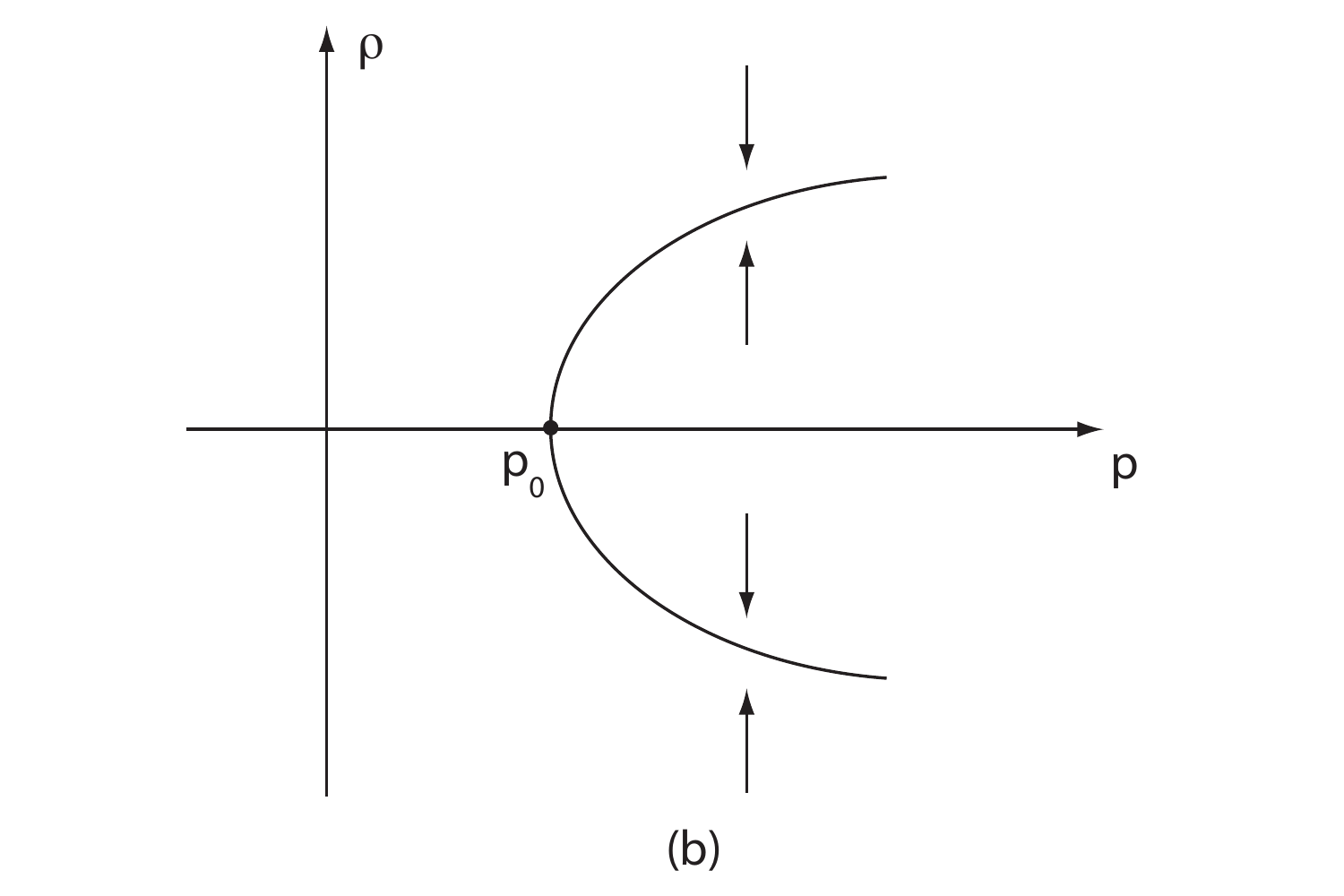}
  \caption{Continuous transition for the case where
$\phi^{\prime}(p_0)>0$}\la{f8.3}
 \end{figure}

\begin{figure}[hbt]
  \centering
  \includegraphics[width=0.35\textwidth]{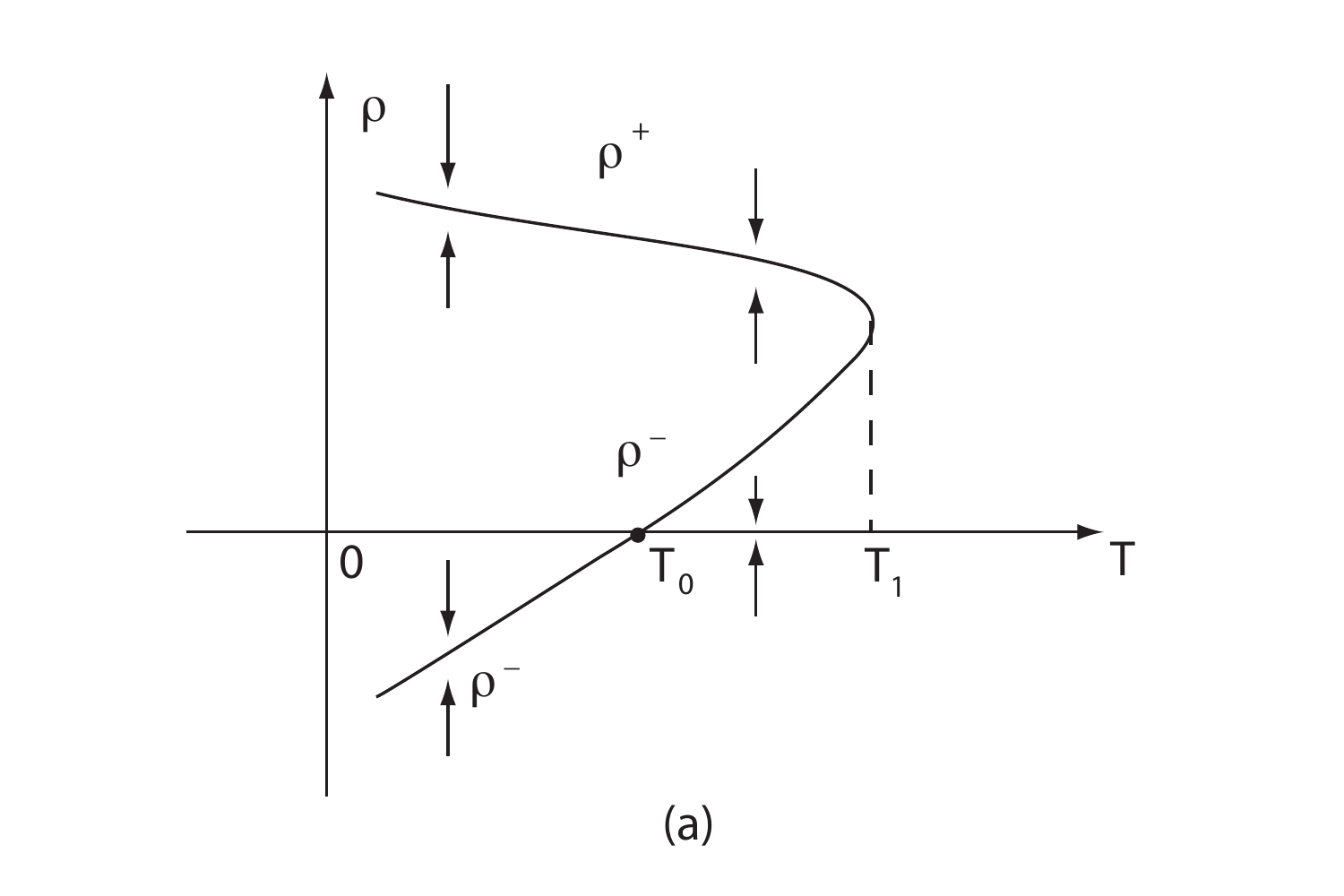} 
  \includegraphics[width=0.35\textwidth]{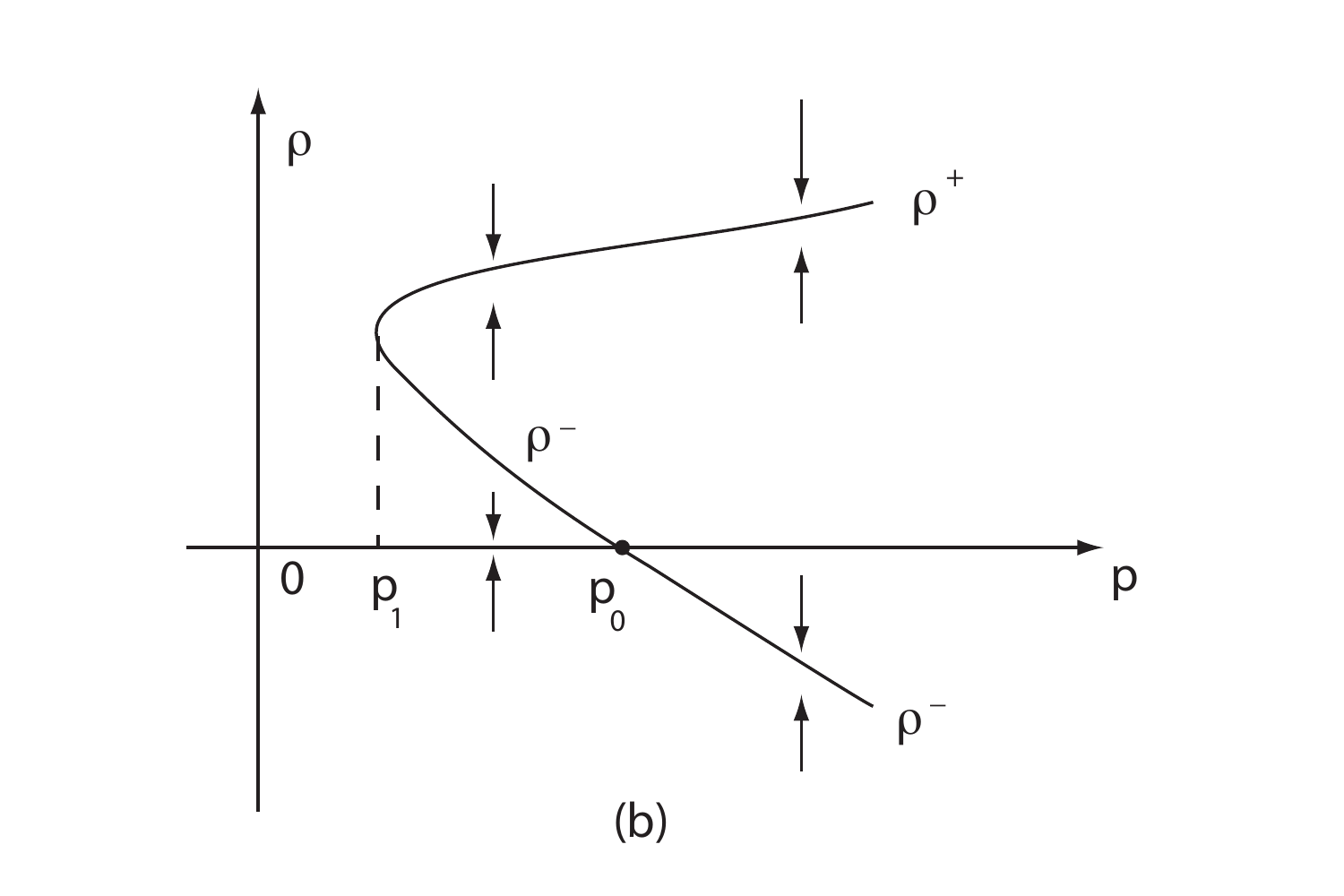}
  \caption{Type-III (mixed) transition for $a_2>0$.}\la{f8.4}
 \end{figure}

\begin{figure}[hbt]
  \centering
  \includegraphics[width=0.35\textwidth]{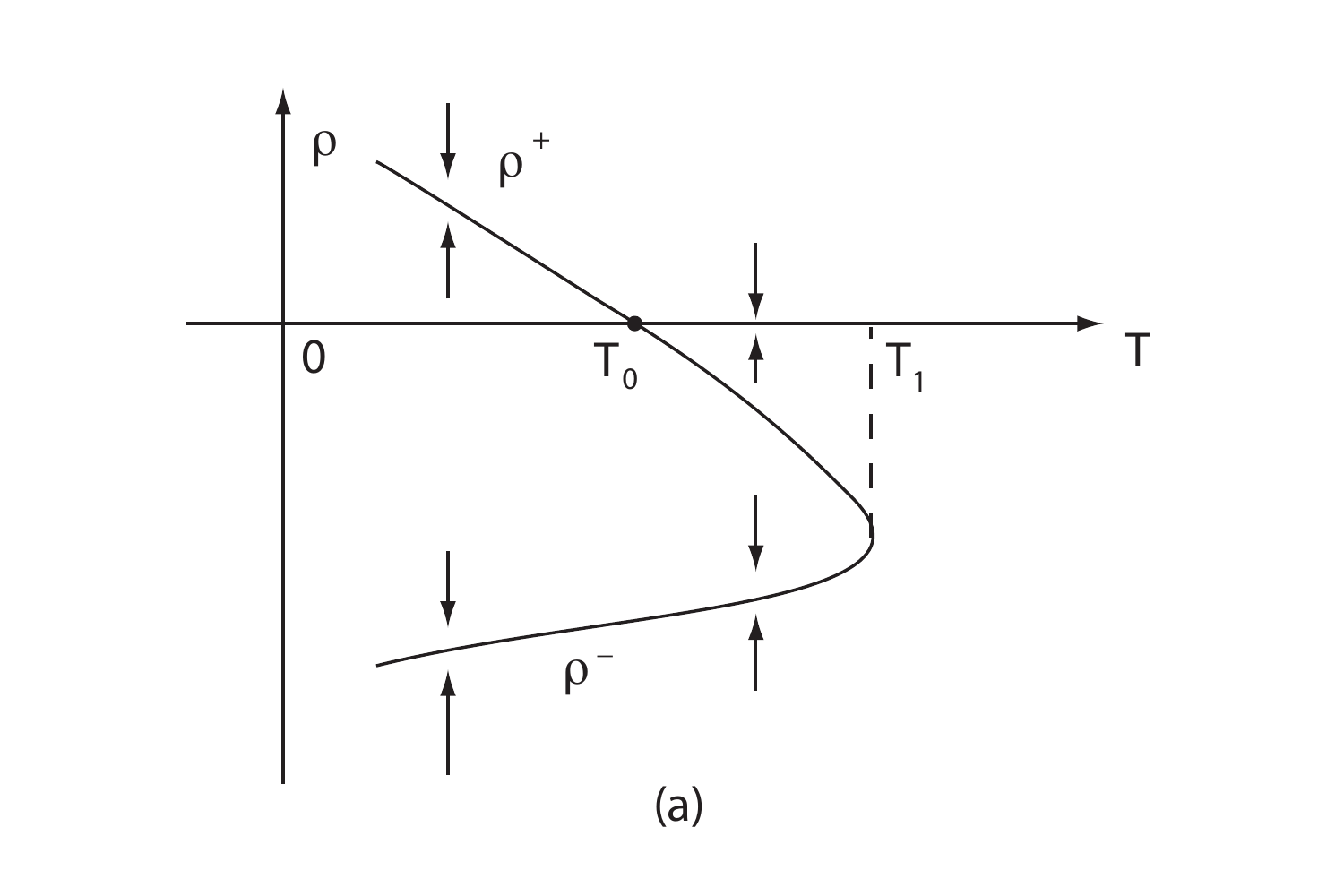}
  \includegraphics[width=0.35\textwidth]{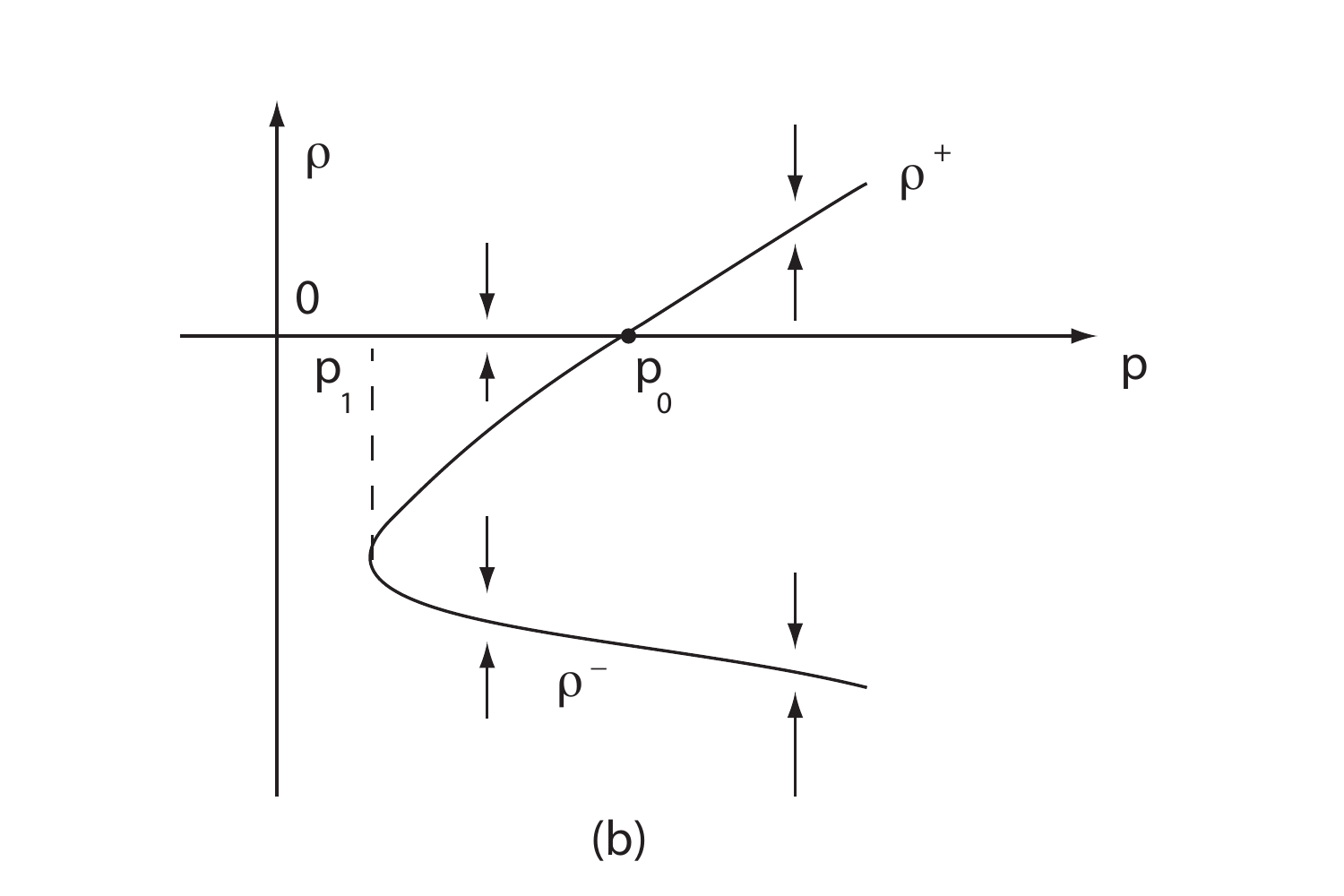}
  \caption{Type-III (mixed) transition for $a_2<0$.}\la{f8.5}
 \end{figure}

\begin{proof}
By (\ref{8.12}), the theorem follows 
from the transition theorems (Theorems \ref{t5.8}  and \ref{t5.9})
and the singularity separation theorem (Theorem \ref{t6.5}). We omit the detailed routine analysis. 
\end{proof}

\section{Physical Conclusions and Predictions}

To discuss the physical significance of Theorem \ref{t8.1}, we recall the
classical $pV$-phase diagram given by Figure \ref{f8.2}. We take $\rho
=1/V$ to replace volume $V$, then the gas-liquid coexistence curve
in   the $\rho$-$P$ plane is illustrated by Figure \ref{f8.6}.
\begin{figure}[hbt]
  \centering
  \includegraphics[width=0.4\textwidth]{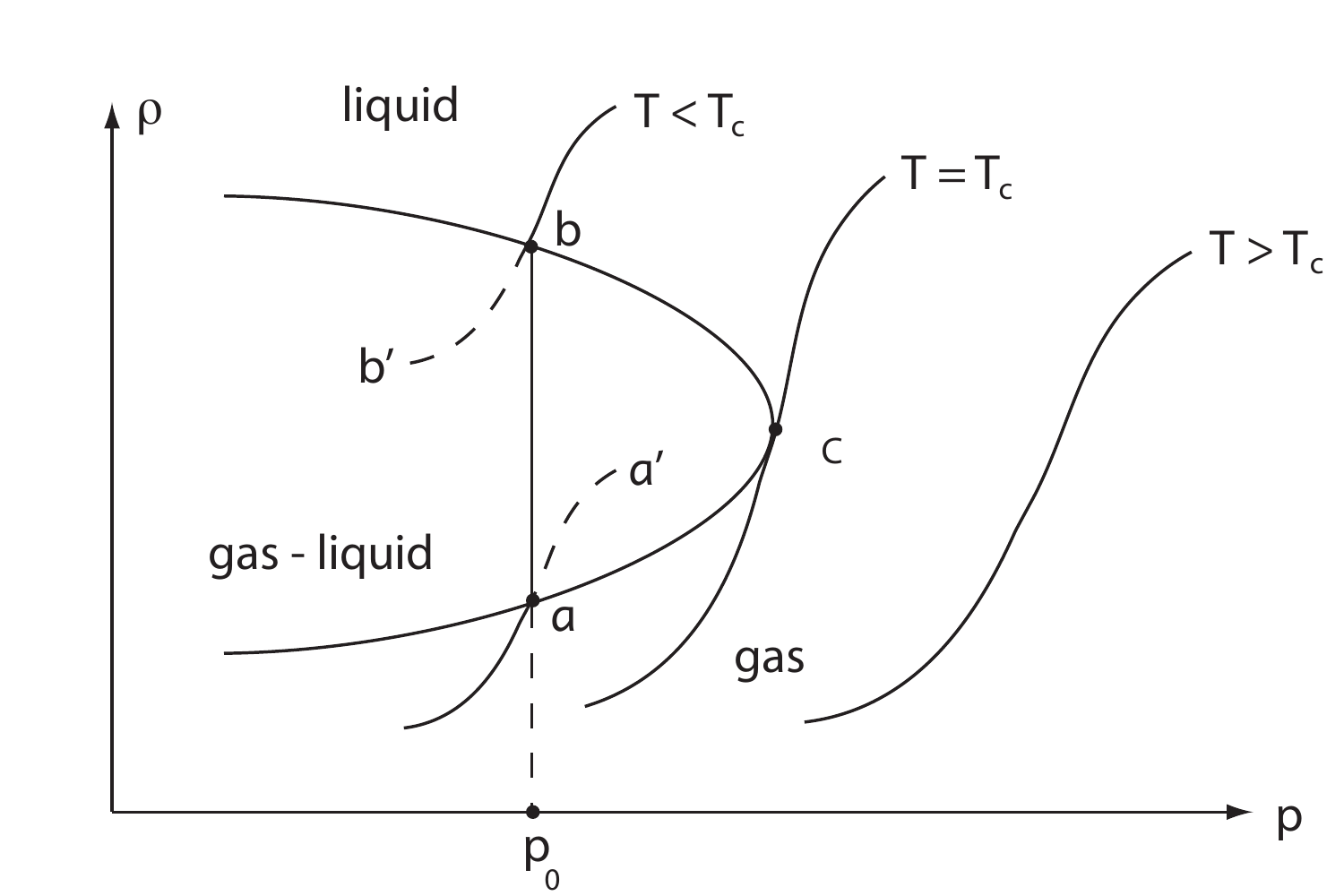}
  \caption{The gas-liquid coexistence curve in $\rho-p$
plane: point $C$ is the Andrews critical point, the dashed lines
$aa^{\prime}$ and $bb^{\prime}$ stand for the metastable states.}
\la{f8.6}
 \end{figure}

According to the physical experiments, in the gas-liquid
coexistence region in the $\rho-p$ phase diagram,  there exist 
metastable states. Mathematically speaking, the metastable states are the
attractors which have a small basin of attraction. In Figure \ref{f8.6},
the dashed lines $aa^{\prime}$ and $bb^{\prime}$ represent the
metastable states, and the points in $aa^{\prime}$ correspond to
super-heated liquid, the points in $bb^{\prime}$ correspond to
super-cooled gas. The $\rho-p$ phase diagram shows that along the
isothermal line $T>T_C$  where $C=(T_C,P_C)$ is the Andrews critical point, 
when the pressure $p$ increases,  the density $\rho$ varies continuously 
from gaseous  to liquid states. However along the isothermal line $T<T_C$
when the pressure $p$ increases to $p_0$ the density $\rho$ will
undergo an abrupt change, and a transition  from gaseous state $a$ to a liquid
state $b$ accompanied with an isothermal exothermal process to
occur. Likewise, such processes also occur in the gas-solid and
liquid-solid transitions.

We now return to discuss Theorem \ref{t8.1}. The steady state solution
$\rho_0=\rho_0(T,p)$ of (\ref{8.11}) can be taken to represent a
desired state in investigating different transition situation. For
example, for studying the gas-liquid transition we can take
$\rho_0$ as the gas density near the transition temperature and
pressure, and for the liquid-solid transition we take $\rho_0$ as
the liquid density, i.e., the lower state density. For convenience,
in the following we alway consider the gas-liquid transition, and
take $\rho_0$ as the gas density.

Let $T_0=\phi (p_0)$, and $\phi^{\prime}(p_0)>0$. Then
Theorem \ref{t8.1} implies the following physical conclusions:

{\sc First},  
near $(T_0,p_0)$,  there are three stable equilibrium
states of (\ref{8.11}) for each of the three cases: $a_2>0$,   $a_2=0$,
or  $a_2<0$:   

\begin{enumerate} 

     \item[a)]  If $a_2<0$,  they are
\begin{equation}
\left.
\begin{aligned} 
& \varphi^+=\rho_0+\rho^+ 
   &&\text{for } |T-T_0|<\delta \text{ \& }  |p-p_0|<\delta,\\
& \varphi^0=\rho_0 &&\text{for }  T_0<T\ \text{or}\ p<p_0,\\
& \varphi^-=\rho_0+\rho^- &&\text{for } T<T_0 \text{ \& }  p>p_0;
\end{aligned}
\right.\label{8.14}
\end{equation}

    \item[b)] If  $a_2=0$,  they are
\begin{equation}
\left.
\begin{aligned}
&  \varphi^{\pm}=\rho_0+\rho^{\pm} &&\text{for } T<T_0   \text{ \& } p>p_0,\\
& \varphi^0=\rho_0 &&\text{for}  T>T_0\ \text{ or } p<p_0,
\end{aligned}
\right.\label{8.15}
\end{equation}
where $\rho^{\pm}=\pm\sqrt{\lambda
/a_3}$;

     \item[c)] If  $a_2<0$,  they are
\begin{equation}
\left.
\begin{aligned} 
& 
\varphi^+=\rho_0+\rho^+ &&\text{for } T<T_0 \text{ \& }  p>p_0,\\
&
\varphi^0=\rho_0 &&\text{for } T>T_0\ \text{or}\ p<p_0,\\
& 
\varphi^-=\rho_0+\rho^- &&\text{for } |T-T_0|<\delta \text{ \& }
|p-p_0|<\delta.
\end{aligned}
\right.\label{8.16}
\end{equation}

\end{enumerate}
Here the state $\varphi^+$ represents the liquid density, $\varphi^0$
the real gas density, and $\varphi^-$ the underlying state
density.

\bigskip

{\sc Second,}
putting $S=\beta^{-1}_1(T-\beta_2\rho^2)$ into
(\ref{8.9}), the free energy becomes
\begin{eqnarray*}
G(\rho
,T,p)&=&-\frac{1}{2}\beta^{-1}_1T^2+\frac{1}{2}(\alpha_1+bp+2\beta^{-1}_1\beta_2T)\rho^2\\
&&-\frac{1}{3}\alpha_2\rho^3+\frac{1}{4}(\alpha_3-2\beta^2_2\beta^{-1}_1)\rho^4-p\rho
.
\end{eqnarray*}
The values of $G$ at equilibrium states of (\ref{8.12}) are given by
\begin{align}
G(\rho_0+\rho^{\pm})=
&G(\rho_0)-\frac{\lambda}{2}
(\rho^{\pm})^2-\frac{a_2}{3}(\rho^{\pm})^3+\frac{a_3}{4}(\rho^{\pm})^4\nonumber\\
=&G(\rho_0)-\frac{\lambda}{4}
(\rho^{\pm})^2-\frac{a_2}{12}(\rho^{\pm})^3.\label{8.17}
\end{align}
By $\lambda =0$ at $(T_0,p_0)$, it follows from
(\ref{8.14})-(\ref{8.17}) that for $(T,p)$ near $(T_0,p_0)$ we
have
\begin{equation}
\left.\begin{aligned}
&  G(\varphi^+)<G(\varphi^0)<G(\varphi^-) &&\text{for}\ a_2>0,\\
&  G(\varphi^+)=G(\varphi^-)<G(\varphi^0) &&\text{for}\ a_2=0,\\
&  G(\varphi^-)<G(\varphi^0)<G(\varphi^+) &&\text{for}\ a_2<0.
\end{aligned}
\right. \label{8.18}
\end{equation}

\bigskip

 {\sc Third, }
 according to Theorem \ref{t8.1} and  the transition diagrams
(Figures \ref{f8.3}-\ref{f8.5}), there are two types of transition behaviors
characterized by some functions $\Phi =\rho (T,p)$ near
$(T,p)=(T_0,p_0)$, which are called the transition functions.
Here, for simplicity we fixed $p=p_0$ and consider $\Phi =\rho
(T)$ as function of $T$. 

If $a_2>0$, the transition functions are
\begin{align*}
&
\Phi^+(T)=\left\{\begin{aligned} 
    &  \rho_0(T) &&   \text{for } T>T^*, \\
    & \rho_0(T)+\rho^+(T)   && \text{for } T<T^*,
  \end{aligned}
    \right.\\
& 
\Phi^-(T)=\left\{\begin{aligned} 
   & \rho_0(T) &&  \text{for } T\geq T_0,\\
    & \rho_0(T)+\rho^- &&  \text{for } T<T_0, 
\end{aligned}
\right.
\end{align*}
for some $T_0\leq T^*<T_1$.
$\Phi^+(T)$ has a finite jump at $T=T^*$ as shown in Figure
\ref{f8.6ab}(a), and $\Phi^-(T)$ is continuous, but has a discontinuous
derivative at $T=T_0$;  see Figure \ref{f8.6ab}(b). 

If $a_2=0$,  the functions
are
\begin{align*}
&
\Phi^+(T)=\left\{\begin{aligned} 
&  \rho_0(T) &&\text{for } T\geq T_0,\\
& \rho_0(T)+\rho^+(T) && \text{for } T<T_0,
\end{aligned}
\right.\\
&
\Phi^-(T)=\left\{\begin{aligned} 
& \rho_0(T) &&   \text{for }T\geq T_0,\\
&\rho_0(T)+\rho^-(T) &&  \text{for }T<T_0,
\end{aligned}
\right.
\end{align*}
and $\Phi^+$ and $\Phi^-$ are continuous with discontinuous
derivatives at $T=T_0$;  see Figure \ref{f8.7}(a) and (b). 

If $a_2<0$,  the  functions are
\begin{align*}
&
\Phi^+(T)=\left\{\begin{aligned} 
& \rho_0(T) && \text{for } T\geq T_0,\\
&\rho_0(T)+\rho^+(T) && \text{for } T<T_0,
\end{aligned}
\right.\\
&
\Phi^-(T)=\left\{\begin{aligned}
& \rho_0(T)  && \text{for } T>T^*,\\
&\rho_0(T)+\rho^-(T) && \text{for } T<T^*,
\end{aligned}
\right.
\end{align*}
for some $T_0\leq T^*<T_1$, 
$\Phi^+$ is continuous with a discontinuous derivative at
$T=T_0$ as shown in Figure \ref{f8.8}(a), and $\Phi^-$ is discontinuous at
$T=T^*$ as shown in Figure \ref{f8.8}(b).

\bigskip

{\sc Fourth,} 
based on physical facts, near the gas-liquid
transition point $(T_0,p_0)$ the density $\rho$ is a decreasing
function of $T$, therefore the transition process $\Phi^+$ is
realistic, and $\Phi^-$ is unrealistic. From the transition
function $\Phi^+(T)$,  the gas-liquid
transition can be well understood.

\begin{figure}[hbt]
  \centering
  \includegraphics[width=0.35\textwidth]{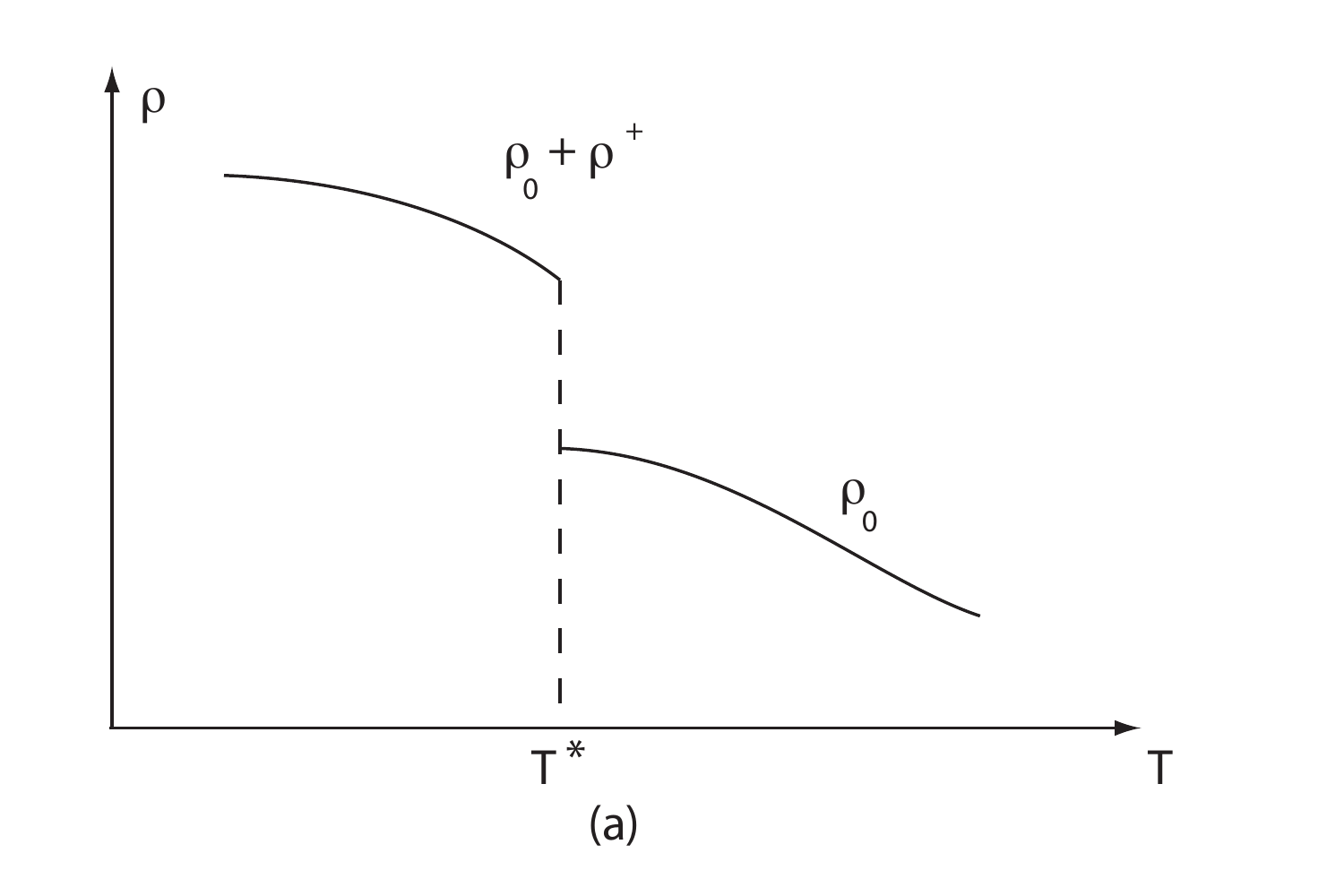}
  \includegraphics[width=0.35\textwidth]{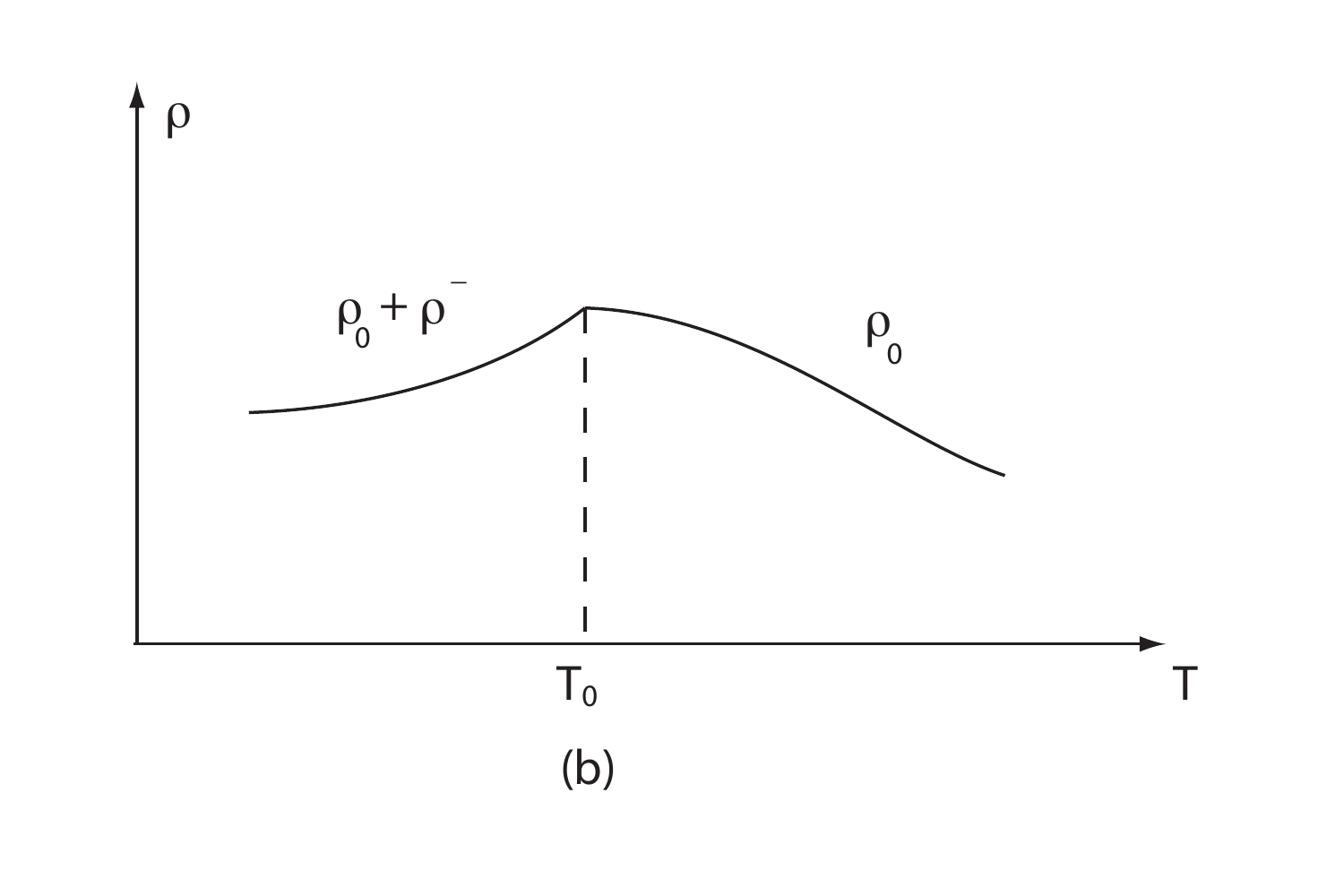}
  \caption{The transition functions for $a_2>0$: (a) the curve
of $\rho =\Phi^+$, (b) the curve of $\rho =\Phi^-$.}\la{f8.6ab}
 \end{figure}

\begin{figure}[hbt]
  \centering
  \includegraphics[width=0.32\textwidth]{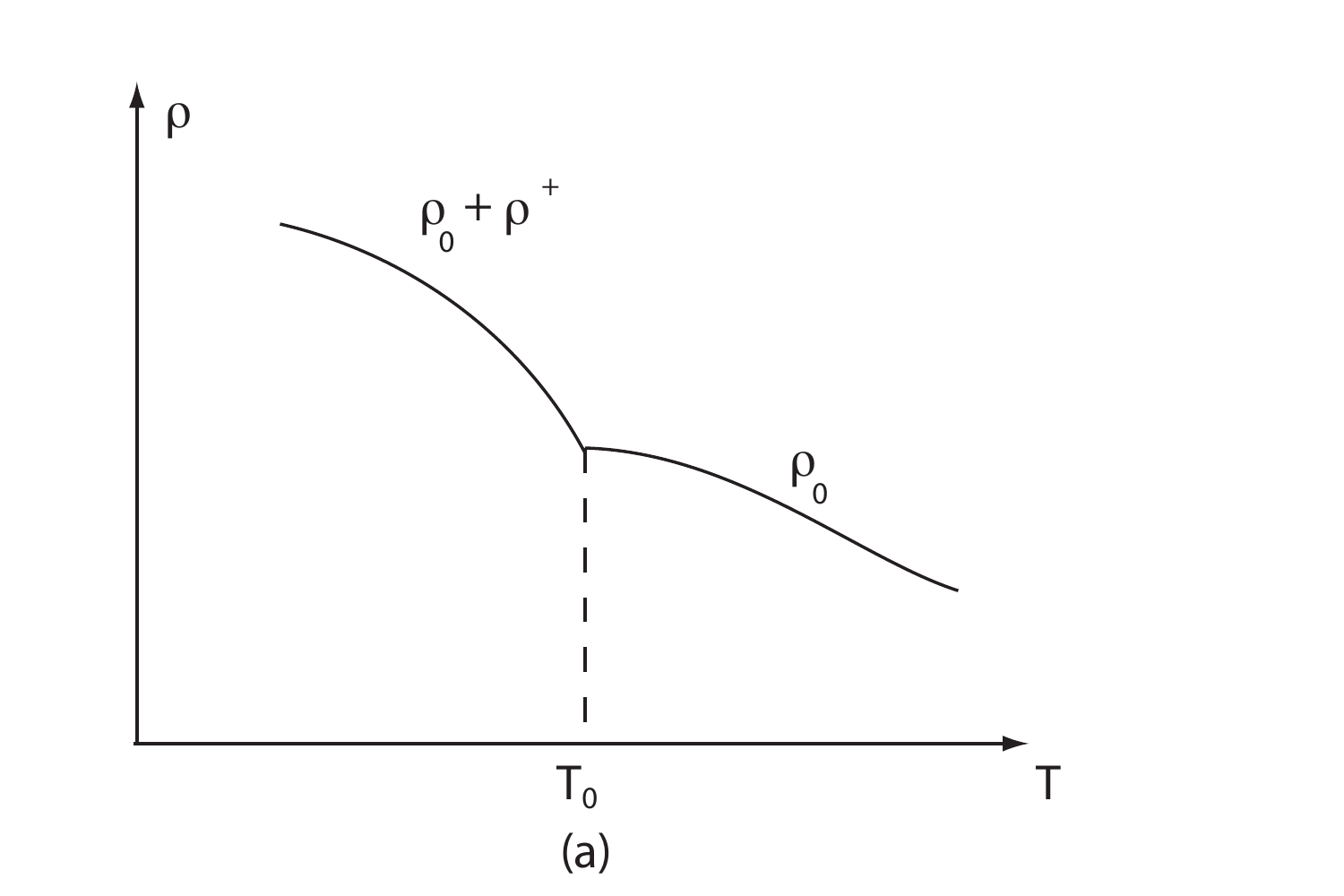}
  \includegraphics[width=0.32\textwidth]{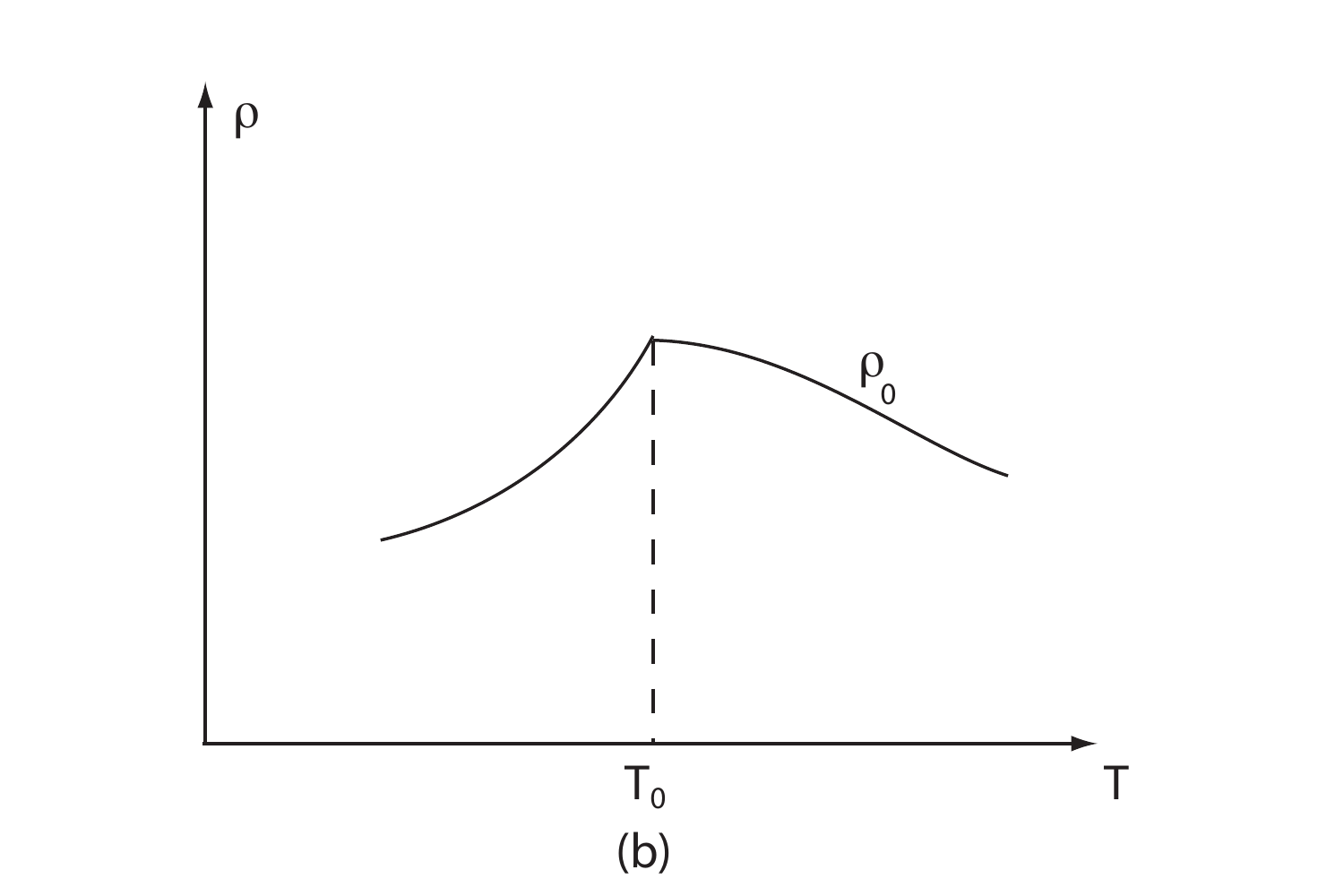}
  \caption{The transition functions for $a_2=0$: (a) the curve
of $\rho =\Phi^+$, (b) the curve of $\rho =\Phi^-$.}\la{f8.7}
 \end{figure}

\begin{figure}[hbt]
  \centering
  \includegraphics[width=0.4\textwidth,height=4cm]{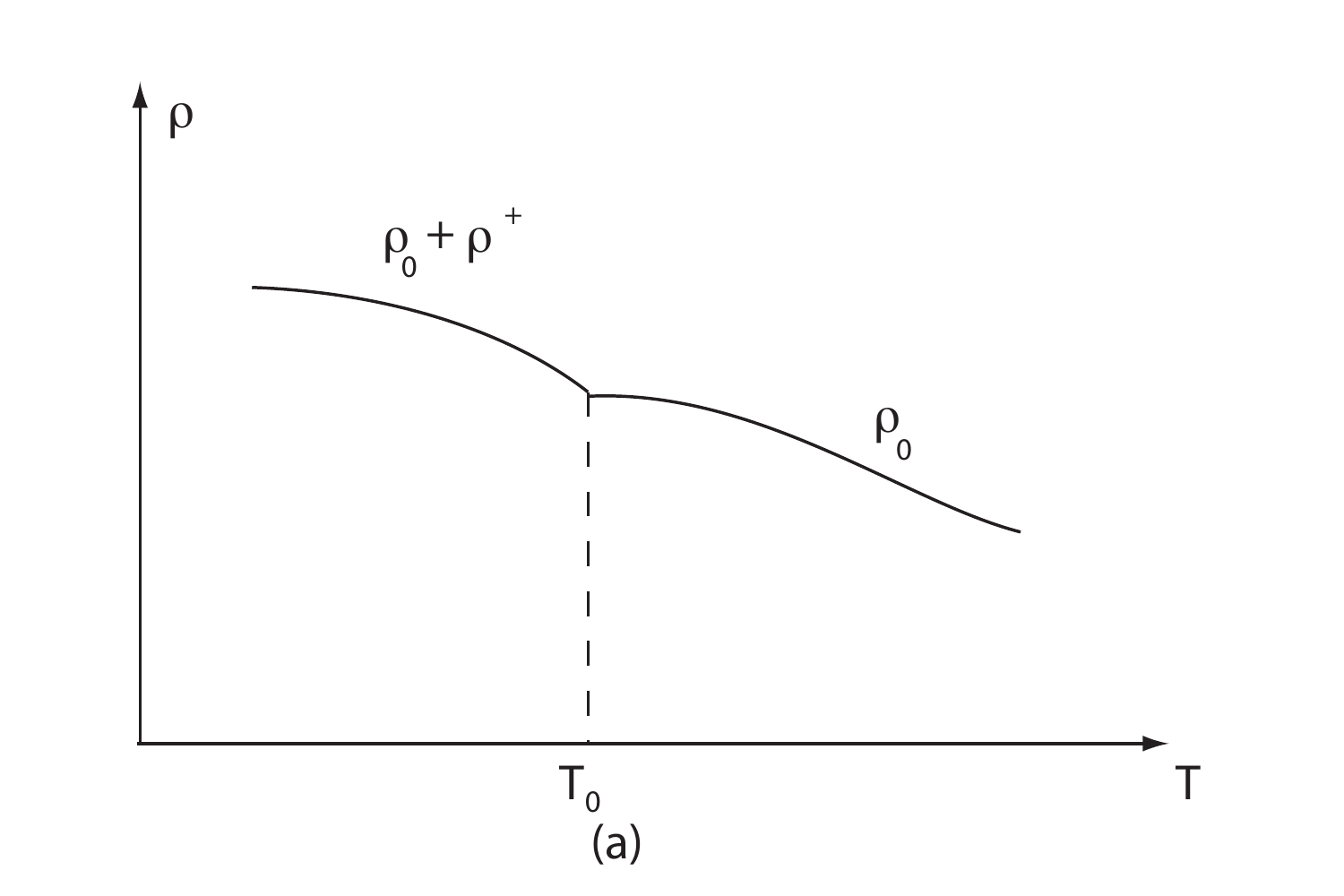}
  \includegraphics[width=0.4\textwidth,height=4cm]{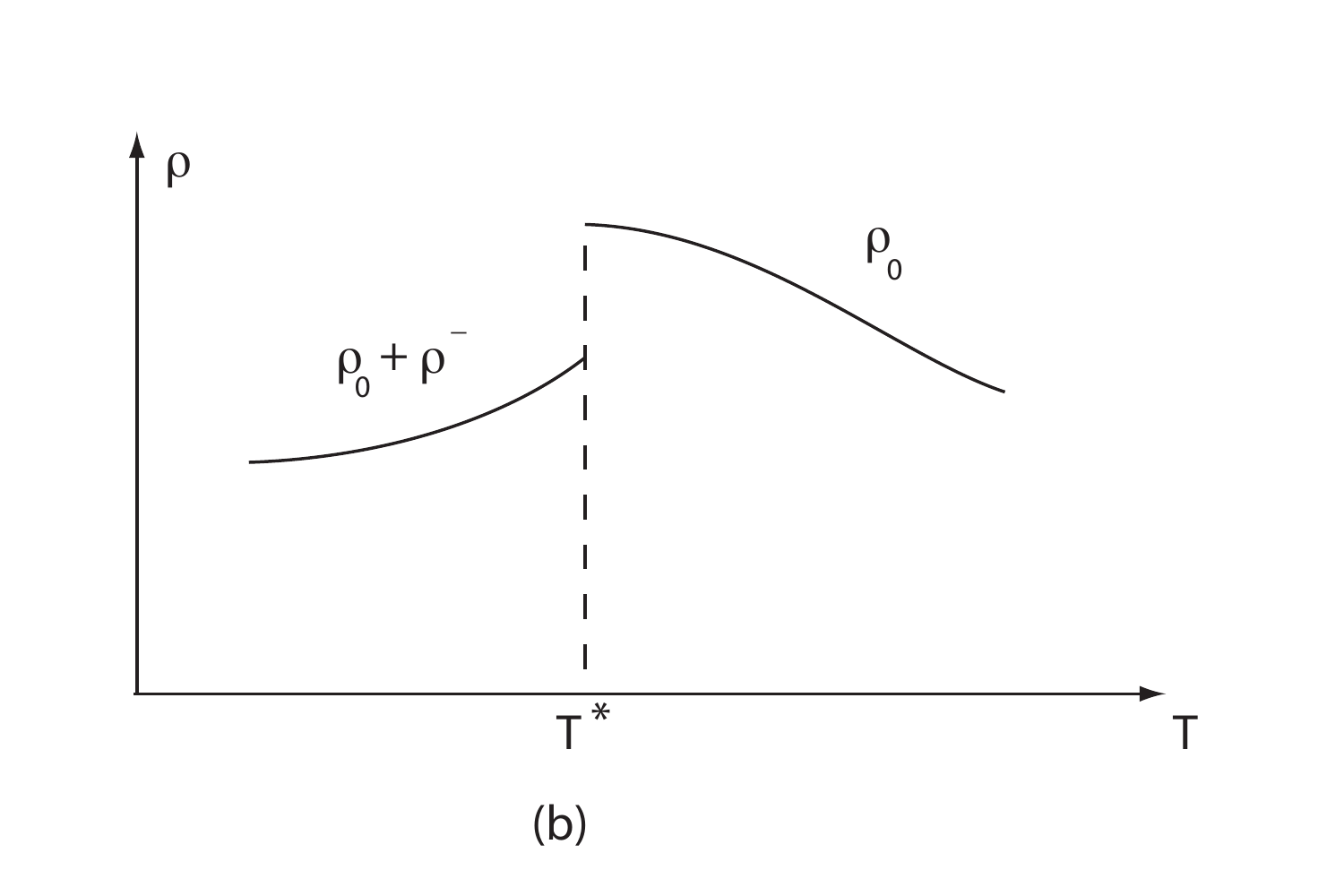}
  \caption{The transition functions for $a_2<0$: (a) the curve
of $\rho =\Phi^+$, (b) the curve of $\rho =\Phi^-$.}\la{f8.8}
 \end{figure}
 
\bigskip

{\sc Fifth, }
 we now explain the gas-liquid transition by using the
transition function $\Phi^+(T)$. Let the two curves
\begin{equation}
\lambda =\lambda (T,p)=0\ \ \ \ \text{and}\ \ \ \
a_2=a_2(T,p)=0\label{8.19}
\end{equation}
intersect at $C=(T_C,p_C)$ in $PT$-plane (see Figure \ref{f8.9}), and the
curve segment $AB$ of $\lambda =0$ is divided into two parts $AC$
and $CB$ by the point $C$ such that
\begin{align*}
& a_2(T,p)>0\ \ \ \ \text{for}\ \ \ \ (T,p)\in AC,\\
& a_2(T,p)<0\ \ \ \ \text{for}\ \ \ \ (T,p)\in CB.
\end{align*}
We shall see that the point $C=(T_C,p_C)$ is the Andrews critical
point, and the curve segment $AC$ is the gas-liquid coexistence
curve, as shown in Figure \ref{f8.1}.
\begin{figure}[hbt]
  \centering
  \includegraphics[width=0.35\textwidth]{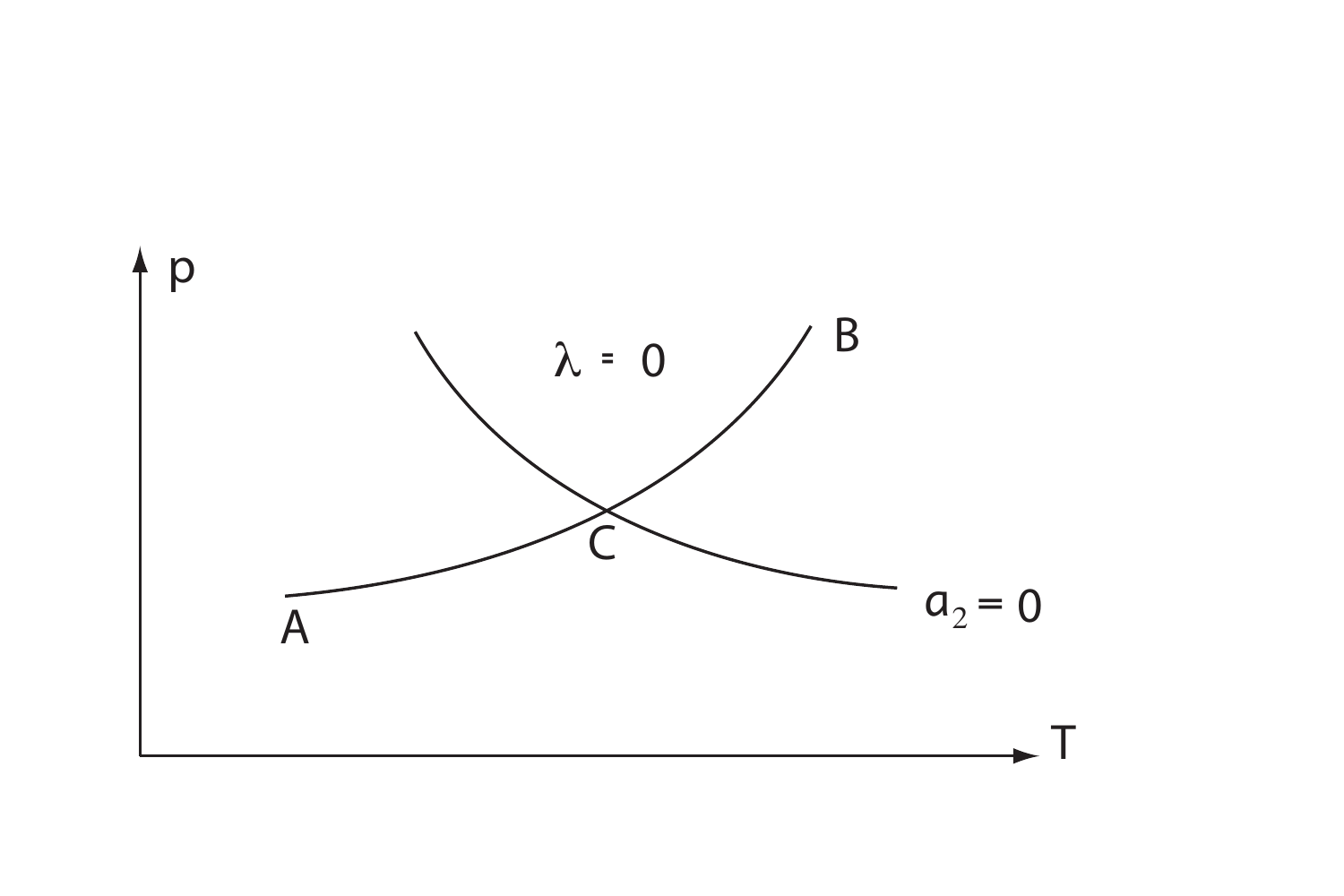}
  \caption{}\la{f8.9}
 \end{figure}

In fact, when $(T_0,p_0)\in AC, a_2(T_0,p_0)>0$ and the transition
function $\Phi^+(T)$ has a jump at $T=T^*$ (see Figure \ref{f8.6}(a)):
$$\rho_0\rightarrow\rho_0+\rho^+,\ \ \ \
\rho^+\simeq\frac{a_2}{a_3}>0.$$ 
Hence  the system undergoes 
a transition from a gaseous state  to a liquid state with an abrupt change in
density. On the other hand, by (\ref{8.18}) there is an energy gap
between the gaseous and liquid states:
$$
\Delta E=G(\rho_0+\rho^+)-G(\rho_0)<0\ \ \ \ \text{at}\ T=T^*.
$$
This energy gap $|\Delta E|$ stands for a latent heat, and $\Delta
E<0$ shows that the transition from a gaseous state to a liquid state is an
isothermal exothermal process, and from a liquid state  to gaseous state is an
isothermal endothermal process.

When $(T_0,p_0)=(T_C,p_C), a_2(T_C,p_C)=0$ and the transition
function $\Phi^+(T)$ is continuous as shown in Figure \ref{f8.7}(a). Near
$T=T_C$,
$$\Phi^+(T)=\rho_0(T)+\sqrt{\lambda /a_3} \qquad \text{for } T<T_C.$$
By the Landau mean field theory, we have
\begin{equation}
\lambda (T)=\alpha (T_C-T)\ \ \ \ (\alpha >0\ \text{a\
constant}).\label{8.20} 
\end{equation} 
Thus  we infer from  (\ref{8.17}) and
(\ref{8.20})   that 
$$
G(\Phi^+(T))=G(\rho_0)-\frac{\alpha^2}{4a_3}(T-T_C)^2\ \ \ \
\text{for}\ T<T_C.
$$ 
The difference of the heat capacity at $T=T_C$ is
$$
\Delta C=-T_C\frac{\partial^2}{\partial
T^2}\left(G(\Phi^+(T_C+0))-G(\rho_0)\right)=\frac{\alpha^2}{2a_3}T_C>0.
$$
Namely the heat capacity has a finite jump at $T=T_C$, therefore
the transition at $T=T_C$ is of  the second order.

When $(T_0,p_0)\in CB, a_2(T_0,p_0)<0$ and the transition function
$\Phi^+(T)$ is continuous as shown in Figure \ref{f8.8}(a). Near $T=T_0$,
$$
\Phi^+(T)=\rho_0(T)+\frac{1}{|a_2|}\lambda (T)+o(|\lambda |)\ \
\ \ T\leq T_0.
$$ 
Notice that (\ref{rhopm}) implies that 
$$\rho^+ = \frac{\lambda}{a_2} + o(|\lambda|).$$
Then we deduce from (\ref{8.17}) and (\ref{8.20})  that if $T>T_0$,
$$G(\Phi^+(T))= G(\rho_0(T)),$$
and if $T\leq T_0$, 
$$G(\Phi^+(T))=G(\rho_0(T))-\frac{\alpha^3}{6|a_2|^2}(T_0-T)^3+o(|T_0-T|^3).$$

Namely, the free energy $G(\Phi^+(T))$ is continuously
differentiable up to the second order at $T=T_0$, and the
transition is of the third order. It implies that as $(T_0,p_0)\in CB$
the Type-I transition at $(T_0,p_0)$ can not be observed by
physical experiments. Therefore we can derive the following
physical conclusion:

\bigskip

\noindent
 {\bf Physical Conclusion V.1.}
{\it The point
$C=(T_C,p_C)$ satisfying (\ref{8.19}) corresponds to the Andrews
critical point, at which the gas-liquid transition is of the second
order. In addition, when $\lambda (T_0,p_0)=0, a_2(T_0,p_0)>0$,
the gas-liquid transition at $(T_0,p_0)$ is of the  first order
accompanied with a latent heat to occur, and when $\lambda
(T_0,p_0)=0, a_2(T_0,p_0)<0$, the transition at $(T_0,p_0)$ is
of the third order.
}

\bigskip

{\sc Sixth,} 
as $\lambda (T_0,p_0)=0$  and $a_2(T_0,p_0)>0$,  the
gas-liquid transition point $(T^*,p^*)$ is in the range $T_0\leq
T^*<T_1$ and $p_1<p^*\leq p_0$; see Figure \ref{f8.4}(a)-(b). In fact, in
the region $T_0\leq T<T_1$ and $p_1<p<p_0$, the two stable states
$\varphi^0=\rho_0(T,p)$ and $\varphi^+=\rho_0(T,p)+\rho^+(T,p)$
are attractors, each possessing a small basin of attraction. Therefore
they correspond to metastable states, and $\varphi^0$ can be
considered as a super cooled gas, while $\varphi^+$ can be considered a
super heated liquid.

\bigskip

{\sc Finally,} 
likewise, we can also discuss the gas-solid and
liquid-solid transitions, and derive the following physical
conclusion:

\bigskip

\noindent
{\bf Physical Conclusion V.2.} 
{\it 
 In the gas-solid and liquid-solid
transitions,  there also exist metastable states. For the gas-solid
case, the metastable states correspond to the superheated solid
and supercooled liquid, and for the liquid-solid case, the
metastable states correspond to the superheated solid and
supercooled liquid.
}

\appendix

\section{Recapitulation of the Dynamic Transition Theory of Nonlinear Systems}
In this appendix we recall some basic elements of the dynamic transition theory developed by the authors \cite{b-book, chinese-book}, which are used to carry out the dynamic transition analysis for the PVT systems in this article. 

Let $X$  and $ X_1$ be two Banach spaces,   and $X_1\subset X$ a compact and
dense inclusion. In this chapter, we always consider the following
nonlinear evolution equations
\begin{equation}
\left. 
\begin{aligned} 
&\frac{du}{dt}=L_{\lambda}u+G(u,\lambda),\\
&u(0)=\varphi ,
\end{aligned}
\right.\label{5.1}
\end{equation}
where $u:[0,\infty )\rightarrow X$ is unknown function,  and 
$\lambda\in \R^1$  is the system parameter.

Assume that $L_{\lambda}:X_1\rightarrow X$ is a parameterized
linear completely continuous field depending contiguously on
$\lambda\in \R^1$, which satisfies
\begin{equation}
\left. 
\begin{aligned} 
&L_{\lambda}=-A+B_{\lambda}   && \text{a sectorial operator},\\
&A:X_1\rightarrow X   && \text{a linear homeomorphism},\\
&B_{\lambda}:X_1\rightarrow X&&  \text{a linear compact  operator}.
\end{aligned}
\right.\label{5.2}
\end{equation}
In this case, we can define the fractional order spaces
$X_{\sigma}$ for $\sigma\in \R^1$. Then we also assume that
$G(\cdot ,\lambda ):X_{\alpha}\rightarrow X$ is $C^r(r\geq 1)$
bounded mapping for some $0\leq\alpha <1$, depending continuously
on $\lambda\in \R^1$, and
\begin{equation}
G(u,\lambda )=o(\|u\|_{X_{\alpha}}),\ \ \ \ \forall\lambda\in
\R^1.\label{5.3}
\end{equation}

Hereafter we always assume the conditions (\ref{5.2}) and
(\ref{5.3}), which represent that the system (\ref{5.1}) has
a dissipative structure.

In the following we introduce the definition of transitions for
(\ref{5.1}).

\bd\la{d5.1}
We say that the system (\ref{5.1}) has a
transition of equilibrium from $(u,\lambda )=(0,\lambda_0)$ on
$\lambda >\lambda_0$ (or $\lambda <\lambda_0)$ if  the following two conditions are 
satisfied:

\begin{itemize}

\item[(1)] when $\lambda <\lambda_0$ (or $\lambda >\lambda_0),
u=0$ is locally asymptotically stable for (\ref{5.1}); and

\item[(2)] when $\lambda >\lambda_0$ (or $\lambda <\lambda_0)$,
there exists a neighborhood $U\subset X$ of $u=0$ independent of
$\lambda$, such that for any $\varphi\in U \setminus \Gamma_{\lambda}$ the
solution $u_{\lambda}(t,\varphi )$ of (\ref{5.1}) satisfies that
$$\left. 
\begin{aligned}
&\limsup_{t\rightarrow\infty}\|u_{\lambda}(t,\varphi
)\|_X\geq\delta (\lambda )>0,\\
&\lim_{\lambda\rightarrow\lambda_0}\delta(\lambda )\geq 0,
\end{aligned}
\right.$$ 
where $\Gamma_{\lambda}$ is the stable manifold of
$u=0$, with  codim $\Gamma_{\lambda}\geq 1$ in $X$
for $\lambda >\lambda_0$ (or $\lambda <\lambda_0)$.
\end{itemize}
\ed

Obviously, the attractor bifurcation of (\ref{5.1}) is a type of
transition. However,  bifurcation and
transition are two different, but related concepts. 
Definition \ref{d5.1} defines the transition of (\ref{5.1}) from a stable
equilibrium point to other states (not necessary equilibrium state).
In general, we can define transitions from one attractor to
another as follows.

\bd\la{d 5.2}
Let $\Sigma_{\lambda}\subset X$ be an
invariant set of (\ref{5.1}). We say that (\ref{5.1}) has a
transition of states from $(\Sigma_{\lambda_0},\lambda_0)$ on
$\lambda >\lambda_0$ (or $\lambda <\lambda_0)$ if the following conditions are 
satisfied:

\begin{itemize}

\item[(1)] when $\lambda <\lambda_0$ (or $\lambda >\lambda_0),
\Sigma_{\lambda}$ is a local minimal attractor, and 

\item[(2)]
when $\lambda >\lambda_0$ (or $\lambda <\lambda_0)$, there exists
a neighborhood $U\subset X$ of $\Sigma_{\lambda}$ independent of
$\lambda$ such that for any $\varphi\in
U \setminus (\Gamma_{\lambda}\cup\Sigma_{\lambda})$, 
the solution $u(t,\varphi)$ of (\ref{5.1}) satisfies that
$$\left. 
\begin{aligned}
&\limsup_{t\rightarrow\infty}\text{\rm dist}(u(t,\varphi),\Sigma_{\lambda})
    \geq\delta (\lambda )>0,\\
&\lim\limits_{\lambda\rightarrow\lambda_0}\delta (\lambda
)=\delta\geq 0,
\end{aligned}
\right.$$ 
where $\Gamma_{\lambda}$ is the stable manifolds of
$\Sigma_{\lambda}$ with codim $\Gamma_{\lambda}\geq 1$.
\end{itemize}
\ed

Let the eigenvalues (counting multiplicity) of $L_{\lambda}$ be given by
$$\{\beta_j(\lambda )\in \C\ \   |\ \ j=1,2,\cdots\}$$
Assume that
\begin{align}
&  \text{Re}\ \beta_i(\lambda )
\left\{ 
 \begin{aligned} 
 &  <0 &&    \text{ if } \lambda  <\lambda_0,\\
& =0 &&      \text{ if } \lambda =\lambda_0,\\
& >0&&     \text{ if } \lambda >\lambda_0,
\end{aligned}
\right.   &&  \forall 1\leq i\leq m,  \label{5.4}\\
&\text{Re}\ \beta_j(\lambda_0)<0 &&  \forall j\geq
m+1.\label{5.5}
\end{align}

The following theorem is a basic principle of transitions from
equilibrium states, which provides sufficient conditions and a basic
classification for transitions of nonlinear dissipative systems.
This theorem is a direct consequence of the center manifold
theorems and the stable manifold theorems; we omit the proof.

\bt\la{t5.1}
 Let the conditions (\ref{5.4}) and
(\ref{5.5}) hold true. Then, the system (\ref{5.1}) must have a
transition from $(u,\lambda )=(0,\lambda_0)$, and there is a
neighborhood $U\subset X$ of $u=0$ such that the transition is one
of the following three types:

\begin{itemize}
\item[(1)] {\sc Continuous Transition}: 
there exists an open and dense set
$\widetilde{U}_{\lambda}\subset U$ such that for any
$\varphi\in\widetilde{U}_{\lambda}$,  the solution
$u_{\lambda}(t,\varphi )$ of (\ref{5.1}) satisfies
$$\lim\limits_{\lambda\rightarrow\lambda_0}\limsup_{t\rightarrow\infty}\|u_{\lambda}(t,\varphi
)\|_X=0.$$ In particular, the attractor bifurcation of (\ref{5.1})
at $(0,\lambda_0)$ is a continuous transition.

\item[(2)] {\sc Jump Transition}: 
for any $\lambda_0<\lambda <\lambda_0+\varepsilon$ with some $\varepsilon >0$, there is an open
and dense set $U_{\lambda}\subset U$ such that 
for any $\varphi\in U_{\lambda}$, 
$$\limsup_{t\rightarrow\infty}\|u_{\lambda}(t,\varphi
)\|_X\geq\delta >0,$$ 
where $\delta >0$ is independent of $\lambda$. 
This type of transition  is also called the discontinuous 
transition. 

\item[(3)] {\sc Mixed Transition}: 
for any $\lambda_0<\lambda <\lambda_0+\varepsilon$  with some $\varepsilon >0$, 
$U$ can be decomposed into two open sets
$U^{\lambda}_1$ and $U^{\lambda}_2$  ($U^{\lambda}_i$ not necessarily
connected):
$$\bar{U}=\bar{U}^{\lambda}_1+\bar{U}^{\lambda}_2,\ \ \
\ U^{\lambda}_1\cap U^{\lambda}_2=\emptyset ,$$ 
such that
\begin{align*}
&\lim\limits_{\lambda\rightarrow\lambda_0}\limsup_{t\rightarrow\infty}\|u(t,\varphi
)\|_X=0   &&   \forall\varphi\in U^{\lambda}_1,\\
& \limsup_{t\rightarrow\infty}\|u(t,\varphi
)\|_X\geq\delta >0 && \forall\varphi\in U^{\lambda}_2.
\end{align*}
\end{itemize}
\et

The following theorem provides sufficient
conditions for continuous transitions and gives local transition
structure.

We consider the transition of (\ref{5.1}) from a simple critical
eigenvalue. Let the eigenvalues $\beta_j(\lambda )$ of
$L_{\lambda}$ satisfy
\begin{equation}
\left. 
\begin{aligned} 
&  
\beta_1(\lambda )
\left\{
 \begin{aligned}
 & <0 &&\text{ if } \lambda <\lambda_0,\\
& =0 &&\text{ if } \lambda =\lambda_0,\\
& >0 &&\text{ if } \lambda >\lambda_0,
\end{aligned}
\right.\\
& \text{Re}\beta_j(\lambda_0)<0 &&\forall j\geq 2,
\end{aligned}
\right.\label{5.35}
\end{equation}
where $\beta_1(\lambda )$ is a real eigenvalue.

Let $e_1(\lambda )$ and $e^*_1(\lambda )$   be  the eigenvectors of
$L_{\lambda}$ and $L^*_{\lambda}$ respectively corresponding to
$\beta_1(\lambda )$ with
$$L_{\lambda_0}e_1=0,\ \ \ \ L^*_{\lambda_0}e^*_1=0,\ \ \ \
<e_1,e^*_1>=1.$$ 
Let $\Phi (x,\lambda )$    be the center manifold
function of (\ref{5.1}) near $\lambda =\lambda_0$. We assume that
\begin{equation}
<G(xe_1+\Phi (x,\lambda_0),\lambda_0),e^*_1>=\alpha
x^k+o(|x|^k),\label{5.36}
\end{equation}
where $k\geq 2$ an integer and $\alpha\neq 0$ a real number.

We have the following transition theorems.
\begin{figure}
  \centering
  \includegraphics[width=0.35\textwidth]{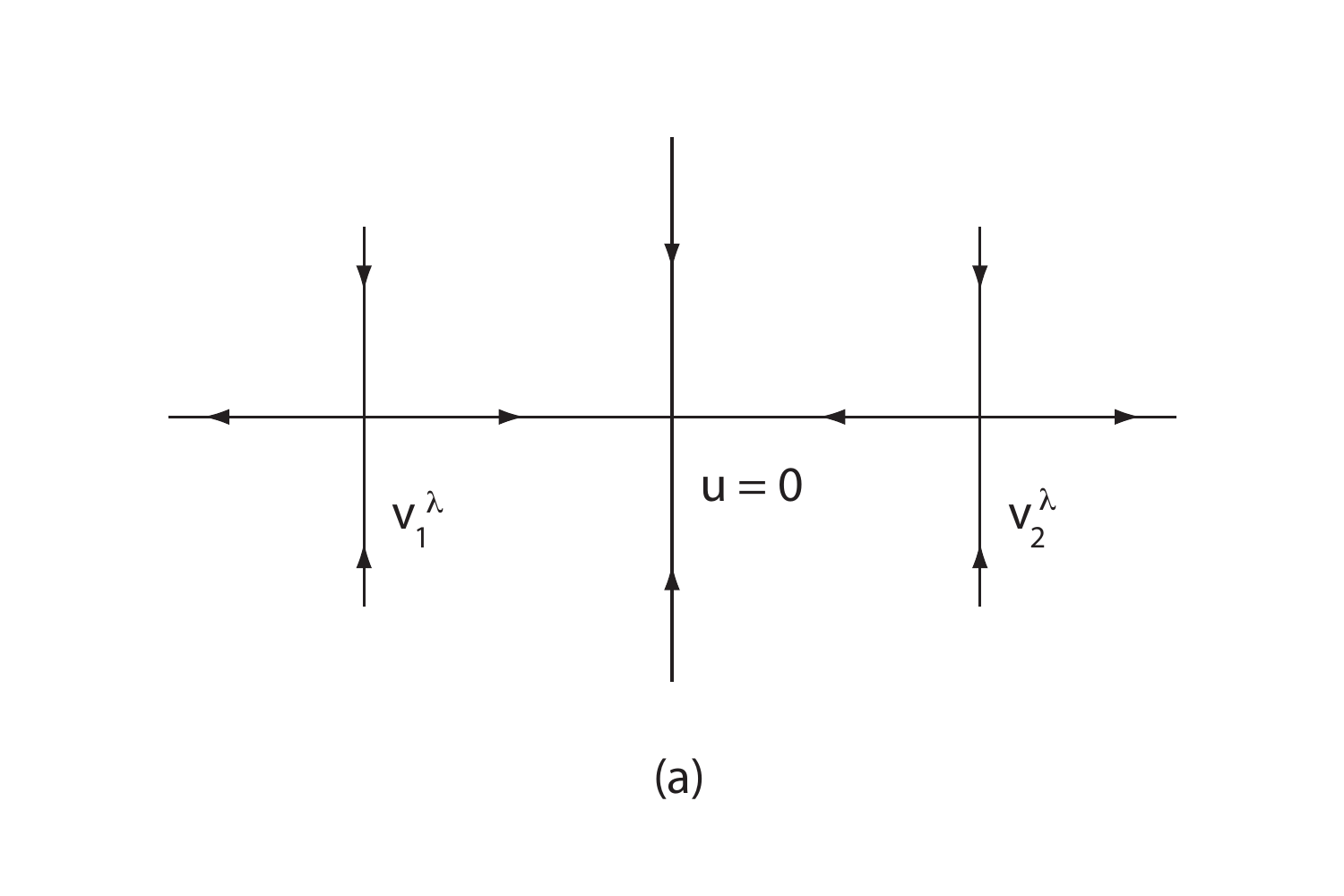} \quad 
  \includegraphics[width=0.2\textwidth]{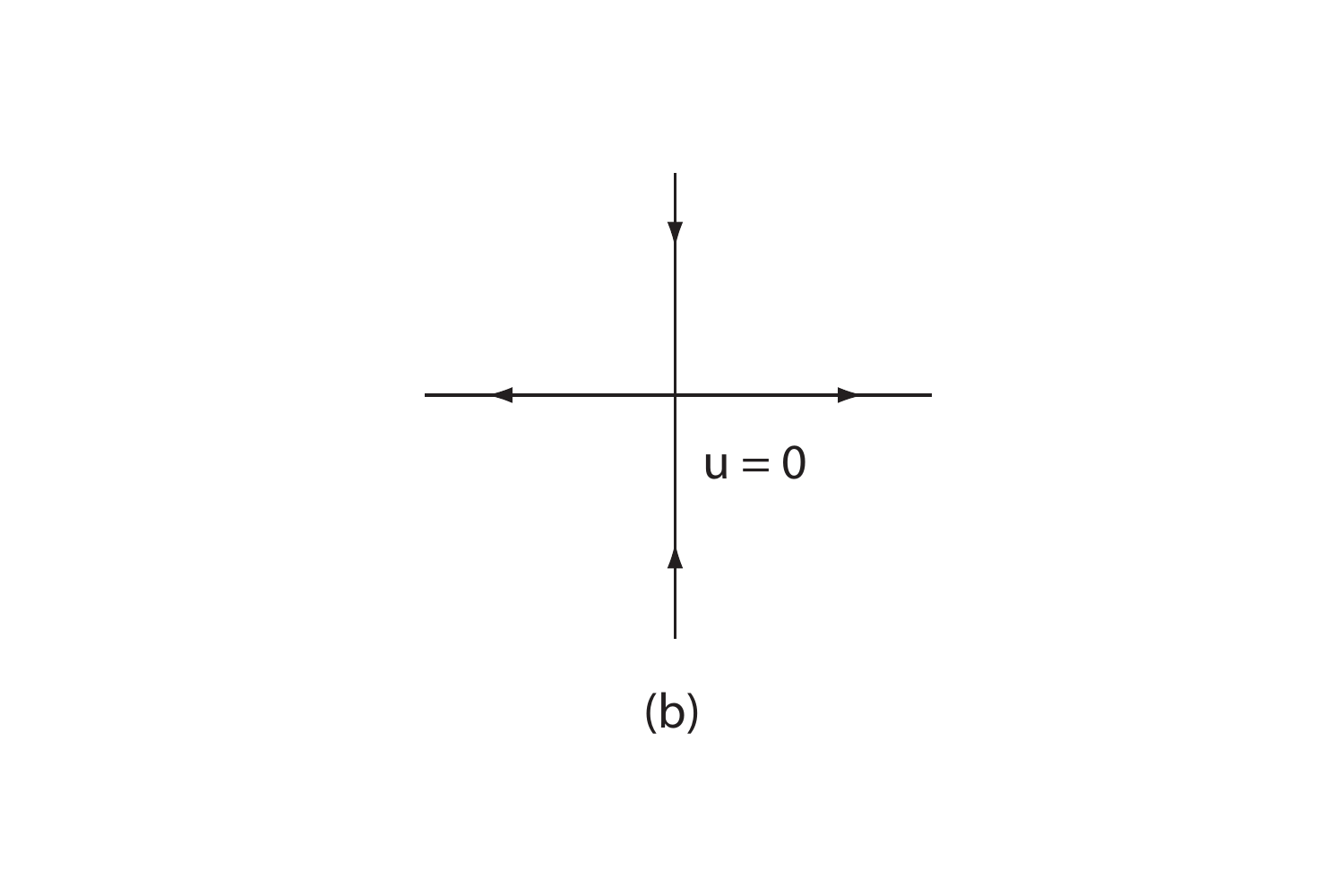}
  \caption{Topological structure of the jump transition of
(\ref{5.1}) when $k$=odd and $\alpha >0$: (a) $\lambda
<\lambda_0$; (b) $\lambda\geq\lambda_0$. Here the horizontal line
represents the center manifold.}\la{f5.5}
 \end{figure}
 \begin{figure}
  \centering
  \includegraphics[width=0.23\textwidth]{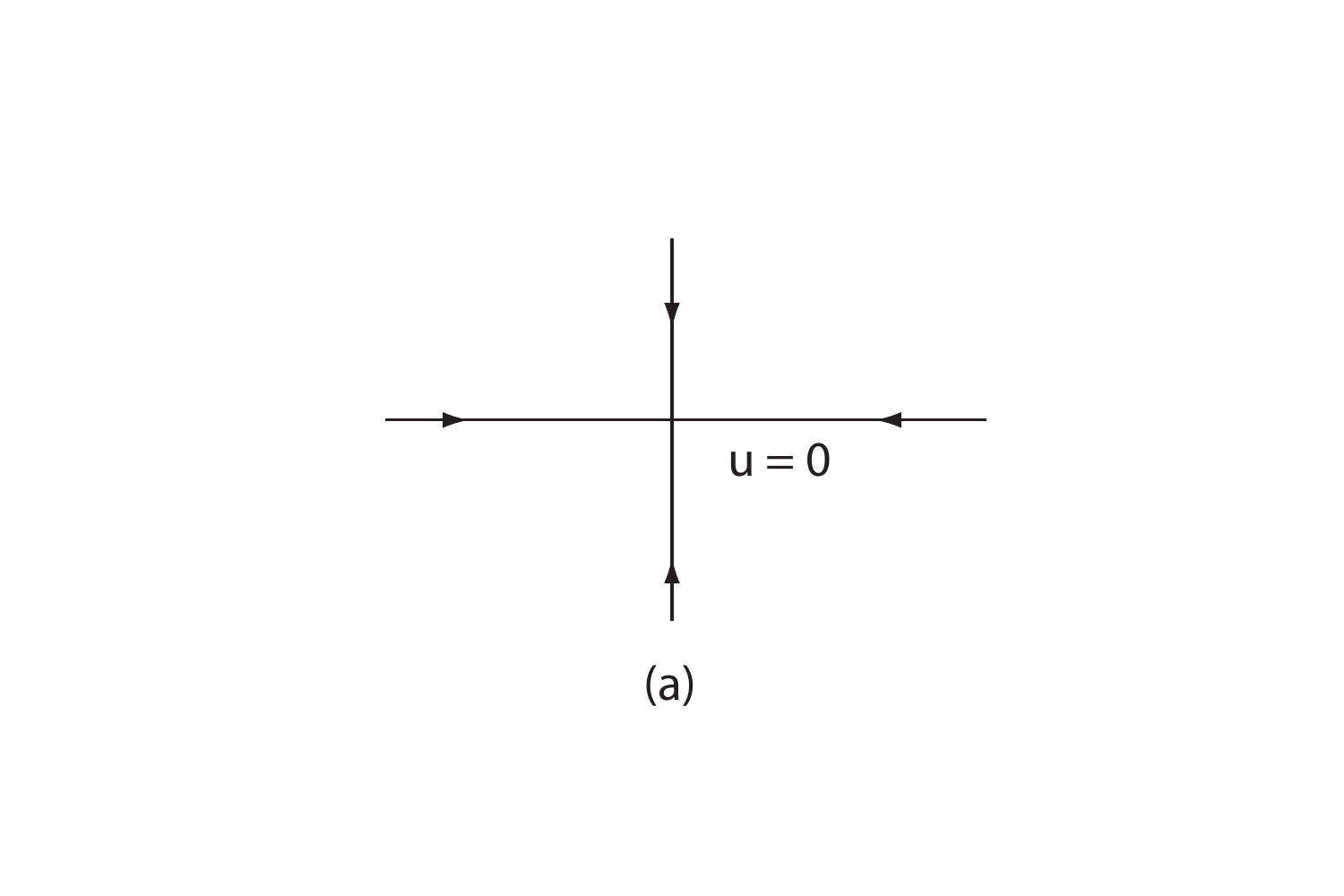}
  \includegraphics[width=0.35\textwidth]{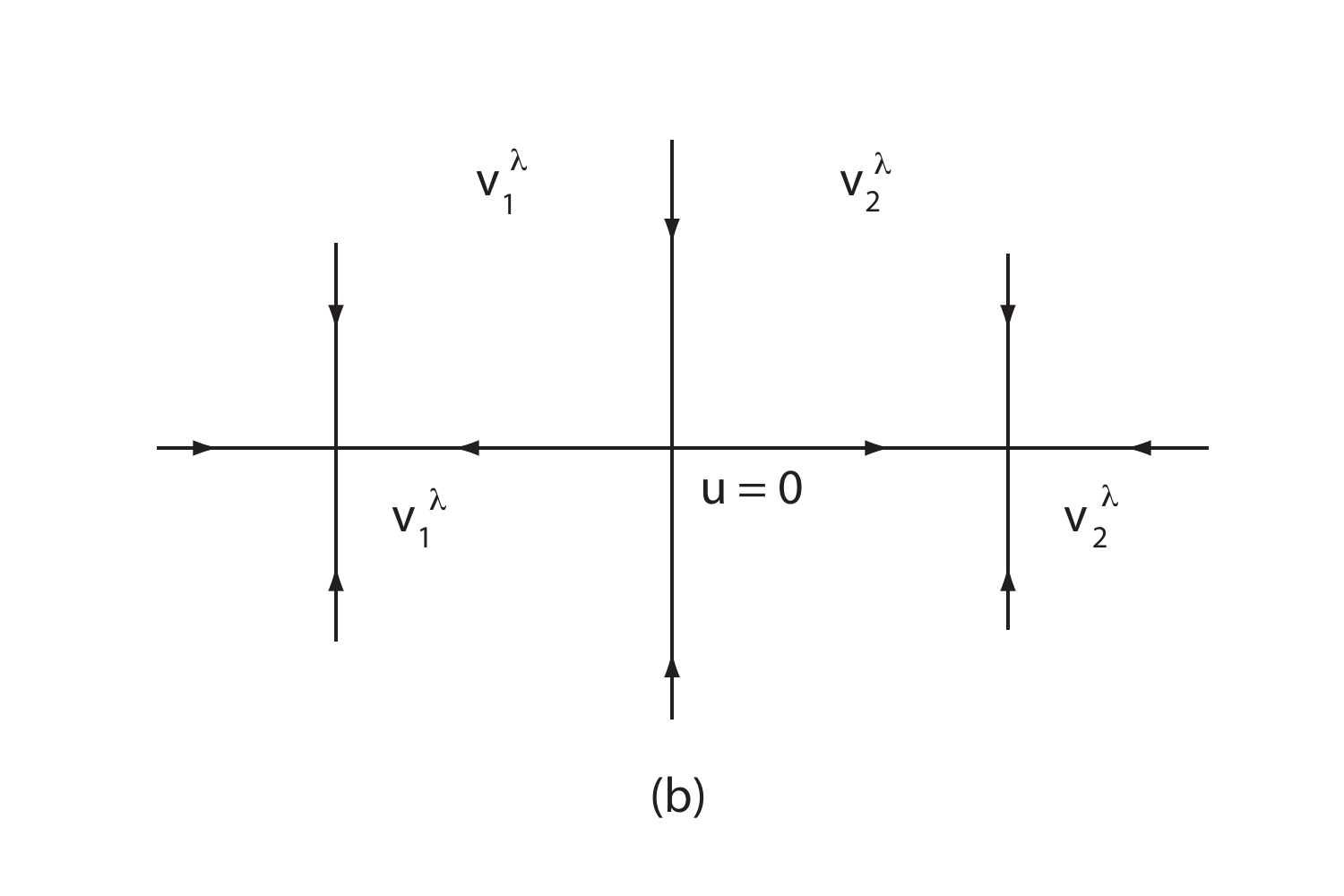}
  \caption{Topological structure of the continuous transition
of (\ref{5.1}) when $k$=odd and $\alpha <0$: (a)
$\lambda\leq\lambda_0$; (b) $\lambda >\lambda_0$.}\la{f5.6}
 \end{figure}
 \begin{figure}
  \centering
  \includegraphics[width=0.32\textwidth]{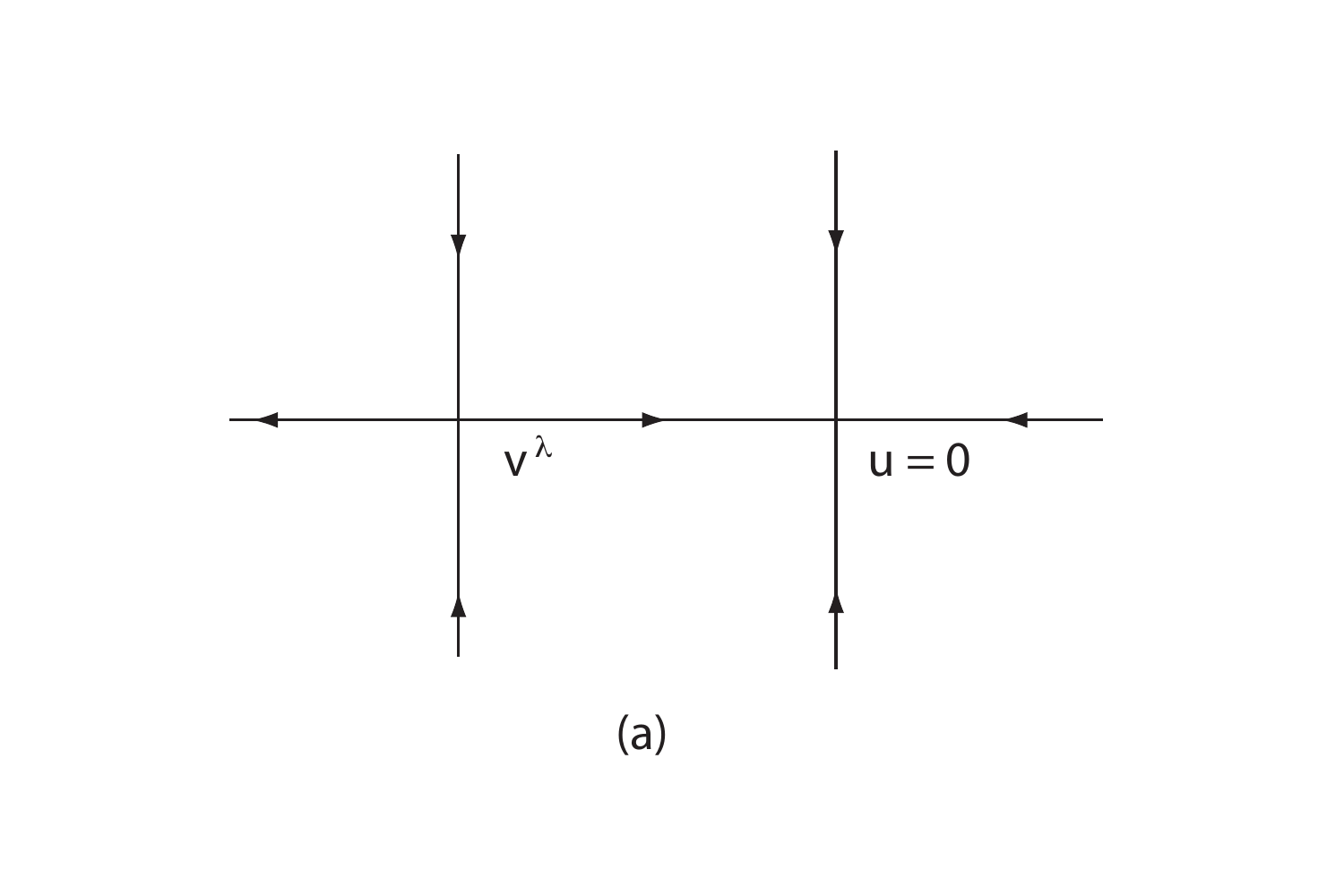}
   \includegraphics[width=0.22\textwidth]{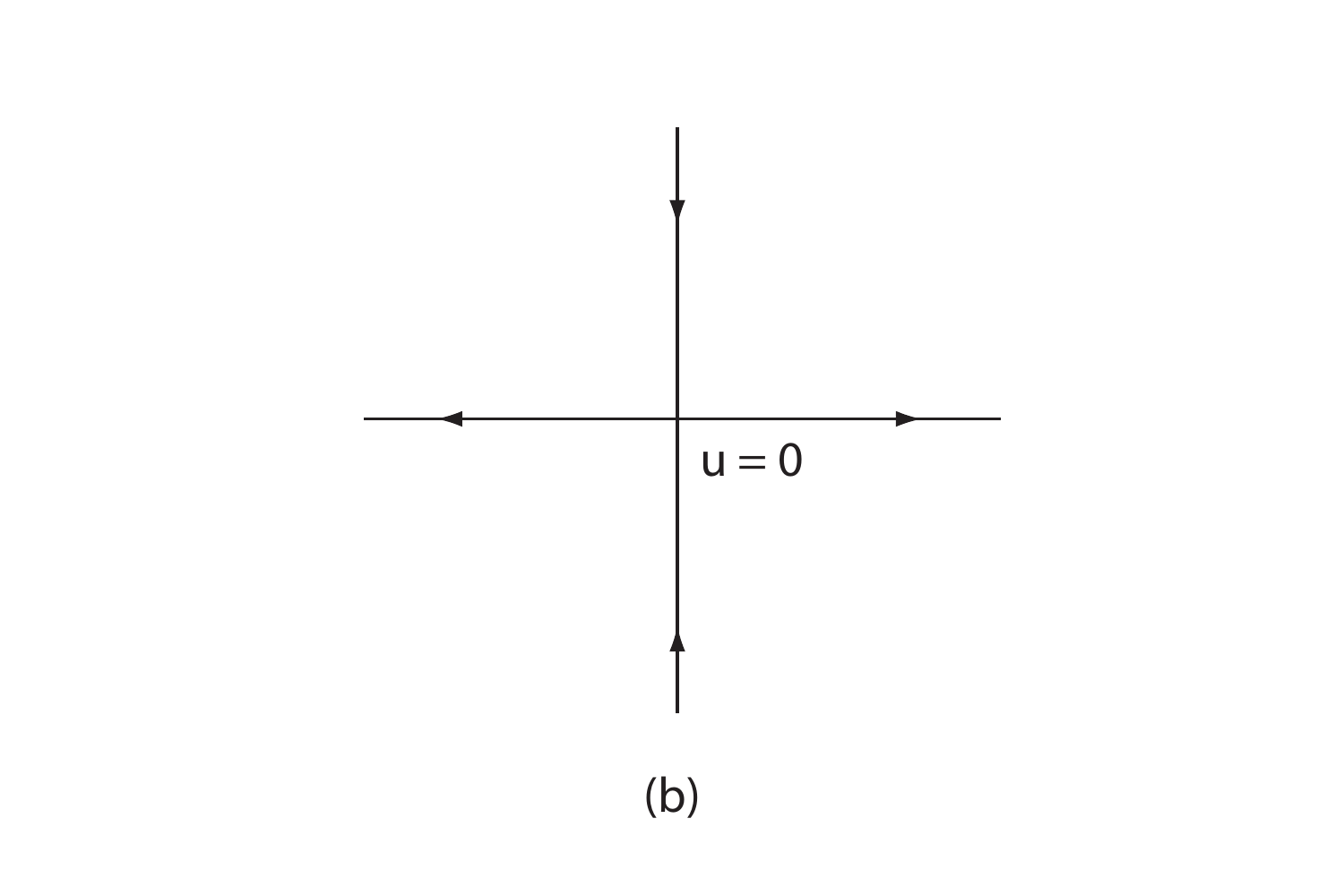} 
   \includegraphics[width=0.32\textwidth]{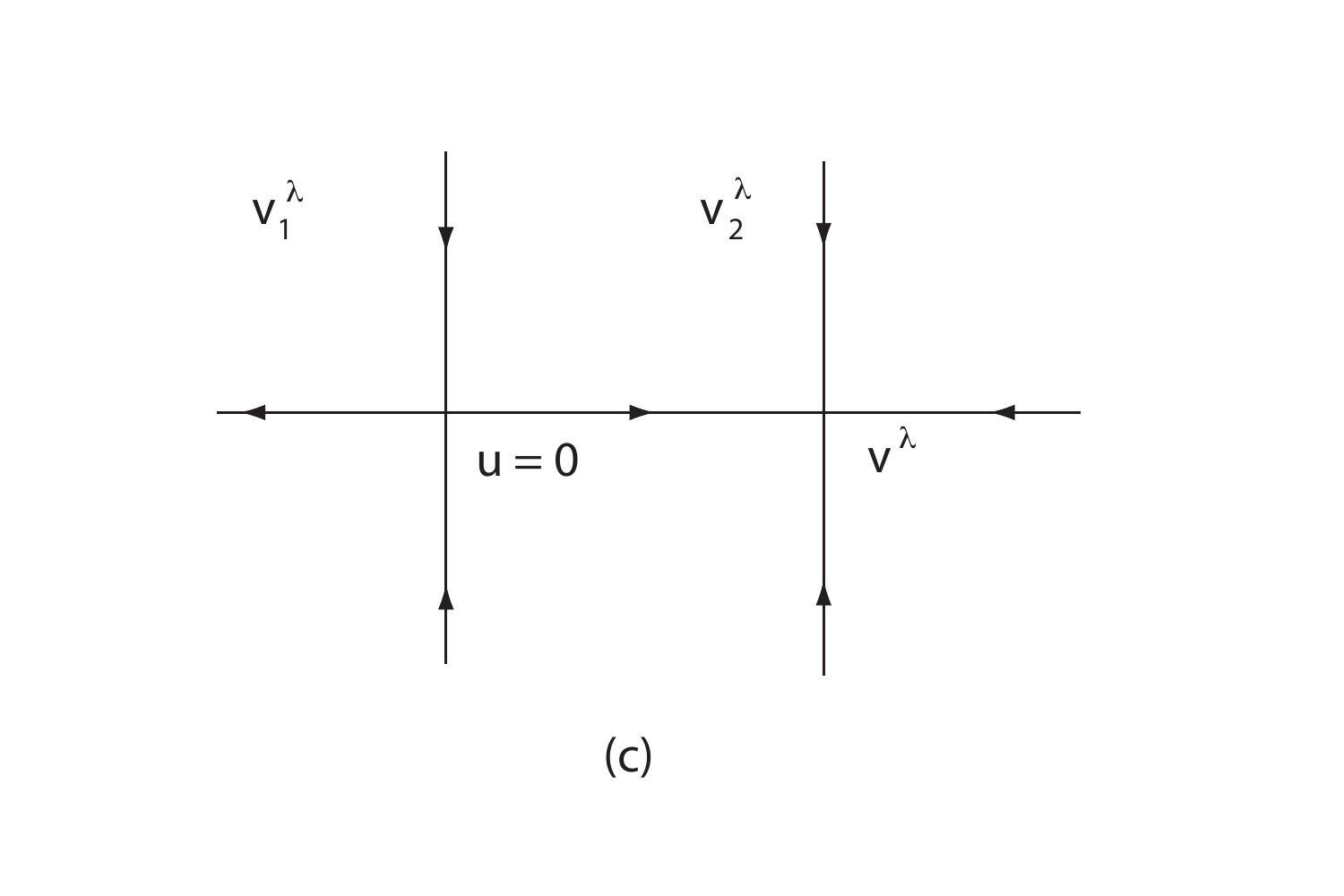}
  \caption{Topological structure of the mixing transition of
(\ref{5.1}) when $k$=even and $\alpha\neq 0$: (a) $\lambda
<\lambda_0$; (b) $\lambda =\lambda_0$; (c) $\lambda >\lambda_0$. Here
$U^{\lambda}_1$ is the unstable domain, and $U^{\lambda}_2$ the
stable domain.}\la{f5.7}
 \end{figure}

\bt\la{t5.8}
Under the conditions (\ref{5.35}) and (\ref{5.36}), 
if $k$=odd and $\alpha\neq 0$ in (\ref{5.36}) then
the following assertions hold true:

\begin{itemize}

\item[(1)] If $\alpha >0$,  then (\ref{5.1}) has a jump
transition from $(0,\lambda_0)$, and bifurcates on $\lambda
<\lambda_0$ to exactly two saddle points $v^{\lambda}_1$ and
$v^{\lambda}_2$ with the Morse index one, as shown in Figure \ref{f5.5}.

\item[(2)] If $\alpha <0$,  then (\ref{5.1}) has a continuous
transition from $(0,\lambda_0)$, which is an attractor bifurcation
 as shown in Figure \ref{f5.6}. 

\item[(3)] The bifurcated singular points $v^{\lambda}_1$ and $v^{\lambda}_2$ 
in the above cases can
be expressed in the following form
$$v^{\lambda}_{1,2}=\pm |\beta_1(\lambda )/\alpha
|^{{1}/{k-1}}e_1(\lambda )+o(|\beta_1|^{{1}/{k-1}}).$$

\end{itemize}
\et

\bt\la{t5.9}
 Under the conditions (\ref{5.35}) and
(\ref{5.36}), if $k$=even and $\alpha\neq 0$, then we have the
following assertions:

\begin{enumerate}

\item (\ref{5.1}) has a mixed transition from
$(0,\lambda_0)$. More precisely, there exists a neighborhood
$U\subset X$ of $u=0$ such that $U$ is separated into two disjoint
open sets $U^{\lambda}_1$ and $U^{\lambda}_2$ by the stable
manifold $\Gamma_{\lambda}$ of $u=0$ satisfying the following properties:

\begin{enumerate}

\item $U=U^{\lambda}_1+U^{\lambda}_2+\Gamma_{\lambda}$,

\item the transition in $U^{\lambda}_1$ is jump, and 

\item the transition in $U^{\lambda}_2$ is
continuous. The local transition structure is as shown in Figure \ref{f5.7}.

\end{enumerate}

\item (\ref{5.1}) bifurcates in $U^{\lambda}_2$ to a unique
singular point $v^{\lambda}$ on $\lambda >\lambda_0$, which is an
attractor such that for any $\varphi\in U^{\lambda}_2$, 
$$\lim\limits_{t\rightarrow\infty}\|u(t,\varphi
)-v^{\lambda}\|_X=0,$$
where $u(t,\varphi )$ is the solution of (\ref{5.1}). 

\item (\ref{5.1})\ bifurcates on $\lambda <\lambda_0$ to a unique saddle
point $v^{\lambda}$ with the Morse index one. 

\item The bifurcated singular point $v^{\lambda}$ can be expressed as
$$v^{\lambda}=-(\beta_1(\lambda )/\alpha
)^{{1}/{(k-1)}}e_1+o(|\beta_1|^{{1}/{(k-1)}}).$$
\end{enumerate}
\et

We consider the equation (\ref{5.1}) defined on the Hilbert spaces
$X=H, X_1=H_1$. Let $L_{\lambda}=-A+\lambda B$. For $L_{\lambda}$
and $G(\cdot ,\lambda ):H_1\rightarrow H$, we assume that
$A:H_1\rightarrow H$ is symmetric, and
\begin{eqnarray}
&&<Au,u>_H\geq c\|u\|^2_{H_{{1}/{2}}},\label{6.46}\\
&&<Bu,u>_H\geq c\|u\|^2_H,\label{6.47}\\
&&<Gu,u>_H\leq -c_1\|u\|^p_H+c_2\|u\|^2_H,\label{6.48}
\end{eqnarray}
where $p>2, c, c_1, c_2>0$ are constants.

\bt\la{t6.5}
 Assume the conditions (\ref{5.3}),
(\ref{5.4}) and (\ref{6.46})-(\ref{6.48}), then (\ref{5.1}) has a
transition at $(u,\lambda )=(0,\lambda_0)$, and the following
assertions hold true:

\begin{enumerate}
\item[(1)] If $u=0$ is an even-order nondegenerate singular point
of $L_{\lambda}+G$ at $\lambda =\lambda_0$, then (\ref{5.1}) has a
singular separation of singular points at some $(u_1,\lambda_1)\in
H\times (-\infty ,\lambda_0)$. 

\item[(2)]  If $m=1$ and $G$
satisfies (\ref{5.36}) with $\alpha >0$ if $k$=odd and $\alpha\neq
0$ if $k$=even, then (\ref{5.1}) has a saddle-node bifurcation at
some singular point $(u_1,\lambda_1)$ with $\lambda_1<\lambda_0$.
\end{enumerate}
\et

\bibliographystyle{siam}

\bibliography{mw-pvt}

\end{document}